\documentclass[12pt,preprint]{aastex}
\def\arcsec {$^{\prime \prime}$}

\def\etal   {{\it et~al.\/}}

\def\HII    {H~{\rm {II}}}
\def\kms    {~km~s$^{-1}$}

\def\OH{$\log({\rm O/H})+12$}
\def\OIII{[{\ion{O}{3}}]}
\def\NIIHa{[\ion{N}{2}]/H$\alpha$}
\def\R23{${\rm R}_{23}$}
\def\OIIIOII{[\ion{O}{3}]/[\ion{O}{2}]}
\def\O32{${\rm O}_{32}$}
\def\Zsun{${\rm Z}_{\odot}$}

\def\numselect {204}

\revised{}
\begin{document}

\title{Metallicities of $0.3<z<1.0$ Galaxies in the GOODS-North Field }

\author{Henry A. Kobulnicky}
\affil{Department of Physics \& Astronomy \\ 
University of Wyoming \\ Laramie, WY 82071 
\\ Electronic Mail: chipk@uwyo.edu}

\author{Lisa J. Kewley}
\affil{Smithsonian Astrophysical Observatory, \\ 
Mail Stop 20, 60 Garden Street, \\ Cambridge, MA 02138
\\ Electronic Mail: lkewley@cfa.harvard.edu}

\author{2004 August 5}

\author{Accepted for Publication in {\it The Astrophysical Journal} }

\vskip 1.cm

\begin{abstract}

We measure nebular oxygen abundances for \numselect\ emission-line
galaxies with redshifts $0.3<z<1.0$ in the Great Observatories Origins
Deep Survey North (GOODS-N) field using spectra from the Team Keck
Redshift Survey (TKRS).  We also provide an updated analytic
prescription for estimating oxygen abundances using the traditional
strong emission line ratio, $R_{23}$, based on the photoionization
models of Kewley \& Dopita (2003).  We include an analytic formula for
very crude metallicity estimates using the
$[N~II]_{\lambda6584}/H\alpha$ ratio.  Oxygen abundances for GOODS-N
galaxies range from $8.2\leq 12+\log(O/H)< 9.1$ corresponding to
metallicities between 0.3 and 2.5 times the solar value.  This sample
of galaxies exhibits a correlation between rest-frame blue luminosity
and gas-phase metallicity (i.e., an L-Z relation), consistent with L-Z
correlations of previously-studied intermediate-redshift samples.  The
zero point of the L-Z relation evolves with redshift in the sense that
galaxies of a given luminosity become more metal poor at higher
redshift. Galaxies in luminosity bins $-18.5<M_B<-21.5$ exhibit a
decrease in average oxygen abundance by 0.14$\pm0.05$ dex from $z=0$
to $z=1$.  This rate of metal enrichment means that $28\pm0.07$\% of
metals in local galaxies have been synthesized since $z=1$, in
reasonable agreement with the predictions based on published star
formation rate densities which show that $\sim38$\% of stars in the
universe have formed during the same interval.  The slope of the L-Z
relation may evolve in the sense that the least luminous galaxies at
each redshift interval become increasingly metal poor compared to more
luminous galaxies. We interpret this change in slope as evidence for
more rapid chemical evolution among the least luminous galaxies
($M_B>-20$), consistent with scenarios whereby the the formation epoch
for less massive galaxies is more recent than for massive galaxies.

\end{abstract}


\section{Metallicity in Galaxy Evolution and Modeling} 

The chemical composition of galaxies is fundamentally important
for tracing galaxy evolution and for modeling galaxy
properties.  Galaxy evolution prescriptions dating from Tinsley (1974,
1980) to present computational codes (e.g., STARBURST99 -- Leitherer
\etal\ 1999; Bruzual \& Charlot 2003; P\'egase -- Fioc, M. \&
Rocca-$\!$Volmerange 1999; GRASIL -- Silva \etal\ 1998) incorporate
metallicity as a primary ingredient in tracking a galaxy's growth.
Metallicity determines a galaxy's UV and optical colors at a given
age, the strength of stellar and interstellar metallic absorption
lines, the shape of its interstellar extinction curve (e.g., Pr\'evot
\etal\ 1984), its dust to gas ratio (Dwek 1998; Issa, MacLaren, \&
Wolfendale 1990), its nucleosynthetic yields (e.g., Woosley \& Weaver 1995),
and perhaps even its rate of star formation (Nishi \& Tashiro 2000).
Overall metal abundance and elemental abundance {\it ratios} trace the
star formation history and nucleosynthetic history of a galaxy
(reviewed by Wheeler, Sneden \& Truran 1989).  These abundances and
ratios also reflect the importance of gas inflow and outflow
in a galaxy's evolution (galactic winds--Matthews \& Baker 1971;
metal-rich galactic winds--Vader 1987) . Observational studies have
now begun to provide detailed optical and infrared measurements of
thousands of galaxies at redshifts from 0.1 to 6 (e.g., Rowan-Robinson
\etal\ 2001; Hippelein \etal\ 2003; Sullivan \etal\ 2004; van Dokkum
\etal\ 2004; Steidel \etal\ 2004, and references therein), spanning
the vast majority of cosmic time.  Successful modeling of the
evolutionary paths of these galaxies will require, among other
parameters, secure measurements of their chemical compositions.

Direct measurements of metallicities\footnote{In this work we focus
primarily on {\it nebular oxygen abundance} as tracer of overall
gas-phase {\it metallicity}, and we use the terms interchangeably.
Other metallicity indicators being explored in distant galaxies
include interstellar absorption lines (Savaglio \etal\ 2004) and
stellar photospheric absorption lines (Mehlert \etal\ 2002).  } in
distant galaxies are now becoming routine, spurred on by larger
telescopes and more capable spectrographs used in deep galaxy surveys.
Kobulnicky \& Zaritsky (1999) first applied classical nebular
diagnostic techniques developed for \HII\ regions in local galaxies to
the global spectra of 14 compact star-forming galaxies at $0.1<z<0.5$.
Their sample spanned a range of galaxy luminosities from $-17>M_B>-22$
and oxygen abundances from $8.25<12+log(O/H)<9.02$\footnote{We take
the solar oxygen abundance to be $12+log(O/H)_\odot\simeq$8.7 based on
the solar oxygen abundance determination of Allende Prieto, Lambert,
\& Asplund (2001)}.  Carollo
\& Lilly (2001) presented oxygen abundances for 15 luminous galaxies
from the Canada-France Redshift Survey (CFRS; Lilly \etal\ 1995) over
the range $0.60<z<0.98$ and $-20.1>M_B>-22.6$.  Both of these early
works concluded that intermediate redshift galaxies were consistent
with the same correlation between luminosity and metallicity (i.e.,
the L-Z relation) observed in local samples (e.g., Faber 1973; Lequeux
\etal\ 1979; Skillman, Kennicutt, \& Hodge 1989; Zaritsky, Kennicutt,
\& Huchra 1994, hereafter ZKH).  Meanwhile, Kobulnicky \& Koo (2000),
Pettini \etal\ (2001), and Shapley \etal\ (2004) used near-infrared
spectroscopy of $2.1<z<3.5$ Lyman break galaxies to measure gas-phase
oxygen abundances.  These authors concluded that the high redshift
objects were 2-4 magnitudes more luminous than $z=0$ galaxies with
comparable $8.3<12+log(O/H)<9$ metallicities and thus were
inconsistent with the local L-Z relation.  Evidence for evolution of
the L-Z relation with epoch, particularly among sub L* galaxies with
$M_B$ fainter than $-20$, grew with metallicity measurements of 64
$0.26<z<0.82$ field galaxies in the Groth Strip Survey
(DGSS--Kobulnicky \etal\ 2003; Ke03) and 66 additional CFRS galaxies
at $0.47<z<0.92$ (Lilly, Carollo \& Stockton 2003--LCS03; Carollo \&
Lilly, 2001--CL01).  Recently, Maier, Meisenheimer \& Hippelein (2004)
measured metallicities for 5 sub L* galaxies at $z\sim 0.4$ and 10 sub
L* galaxies at $z\sim0.64$ from the CADIS emission-line survey.  These
additional sub L* galaxies provide further support for the evolution
of the L-Z relation with epoch.

In this paper we present gas-phase oxygen abundance measurements for
204 emission-line galaxies from $0.3<z<0.93$ in the Great
Observatories Origins Deep Survey-North (GOODS-North; Dickinson \etal\
2001) field using the publicly available spectra obtained as part of
the Team Keck Treasury Redshift Survey (TKRS; Wirth \etal\ 2004).  The
new data double the number of metallicities previously available for
this redshift range and constitute the highest quality spectra yet
available for chemical analysis.  In addition to providing new
constraints on the chemical enrichment of galaxies over the last 8
Gyr, it is our hope that these measurements will prove useful for the
community in modeling the evolution of galaxies in this well-studied
cosmological field.  We combine these new TKRS data with existing
emission line measurements from the literature (CFRS--LCS03 ;
DGSS--Kobulnicky \etal\ 2003) to assess the chemo-luminous evolution
of star-forming galaxies out to $z=1$. Where applicable, we adopt a
cosmology with $H_0$=70 \kms\ Mpc$^{-1}$, $\Omega_m=0.3$, and
$\Omega_\Lambda=0.7$.

\section{Data Analysis}

\subsection{Target Selection}

The TKRS consists of Keck telescope spectroscopy with the DEIMOS
(Faber \etal\ 2003) spectrograph over the nominal wavelength range
4600 \AA\ -- 9800
\AA.  Targets include both stars and galaxies, 
selected in an unbiased manner from all objects with $R\leq24.4$ in
ground-based optical images (Wirth \etal\ 2004).  Slitmasks with
1\arcsec\ wide slits tilted to align with a galaxy's major axis
provided up to 100 spectra per 1200 s exposure.  Spectra have typical
resolutions of $\sim$3.5 \AA\ and total integration times of 3600 s.

We searched the publicly available TKRS spectroscopic database for
galaxies with nebular emission lines suitable for chemical
analysis.  Only galaxies where it was possible to measure all of the
requisite [O~II]$\lambda3727$, H$\beta$, and [O~III]$\lambda$5007
lines were retained.  These criteria necessarily exclude objects at
redshifts of $z\lesssim0.3$ since the requisite [O~II]$\lambda$3727
line falls below the blue limit of the spectroscopic setup.  Likewise,
objects with redshifts $z\gtrsim 1.0$ are excluded because the
[O~III]$\lambda$5007 line falls beyond the red wavelength limit of the
survey.  In the 2004 February public release of the TKRS there were
1536 objects with secure redshifts.  Of these, 1044 fell within our
redshift limits.  Of the 1044 candidates, 497 were removed from
consideration because no emission lines were present in the spectrum.
Next, 94 galaxies were removed from the sample because one of the
requisite strong emission lines fell off the end of the wavelength
coverage (or in between the red and blue spectral regions) due to the
object's placement on the slitmask.  Following Kobulnicky \etal\
2003), we removed 94 objects in the redshift ranges $0.410<z<0.426$,
$0.52<z<0.54$, and $0.56<z<0.58$. For these intervals, atmospheric
$O_2$ absorption troughs between 6865 \AA\ -- 6920 \AA\ (the ``B
band'') and between 7585 \AA\ -- $\sim7680$ \AA\ (the ``A band'')
prohibit accurate measurement of emission lines. Another 122 objects
were removed from the sample because the $H\beta$ emission line was
absent or too weak (S/N$<$8:1) for reliable chemical determinations
(see Kobulnicky, Kennicutt, \& Pizagno 1999 for a discussion of errors and
uncertainties).  The spectra of objects rejected due to a weak
$H\beta$ line are usually dominated by stellar continuum rather than
nebular emission from star-forming regions.  Most local early-type
spirals and elliptical galaxies share these spectral characteristics.
For these objects, H$\beta$ is seen in absorption against the stellar
spectrum of the galaxy.  Thus early type galaxies with older stellar
populations are preferentially rejected in favor of late type galaxies
with larger star formation rates. An additional 31 galaxies had to be
rejected because some combination of strong night sky emission bands,
low continuum, or poor continuum subtraction caused the extracted 1-D
spectrum to have continuum levels that were either negative or very
close to zero.  Because chemical analysis requires emission lines 
powered by normal stellar ionizing radiation fields (as opposed to
nonthermal sources from active galactic nuclei), we rejected 9 objects
exhibiting canonical AGN or LINER signatures: broad emission lines (2
objects), or $EW_{[Ne~III]\lambda3826}/EW_{[O~II]\lambda3727}$ ratios
exceeding 0.4 (7 objects) (e.g., Osterbrock 1989).  The resulting
usable sample contained \numselect\ emission-line galaxies.

Twenty-seven of the \numselect\ selected galaxies have measurable
$H\beta$ but immeasurably weak ($S/N<3:1$) [O~III] lines.  In
principle, such objects should be included in the sample to avoid
introducing a metallicity bias, but it is not possible to compute
reliable metallicities if the oxygen lines are not detected with S/N
ratio of 8:1 or better. We have retained these objects in the sample
and measure upper limits on the [O~III] line strengths which
are used below to compute lower limits on oxygen abundances.

The 204 galaxies in our sample appear in Table~\ref{src.tab}, along
with their TKRS identifications from Wirth \etal\ (2004) in column 2,
their GOODS-N\footnote{The GOODS-N data descriptions, photometric
catalogs, and publicly available data can be found at
http://www.stsci.edu/science/goods.  } designations in column 3,
spectroscopic redshifts in column 4, TKRS R-band magnitude in column 5,
and  GOOD-N ``i'' (F775W), photometry
in column 6.  From the observed
magnitudes and and known redshifts, we computed\footnote{We are
grateful to C. Willmer of the DEEP2 Redshift Survey Team (Davis \etal\
2003) for computing the K-corrections.  Details are presented in
Willmer \etal\ (2004).}  restframe absolute blue magnitudes, $M_B$,
and colors, $U-B$, which appear in columns 7 and 8 of
Table~\ref{src.tab}.

\begin{figure}
\plotone{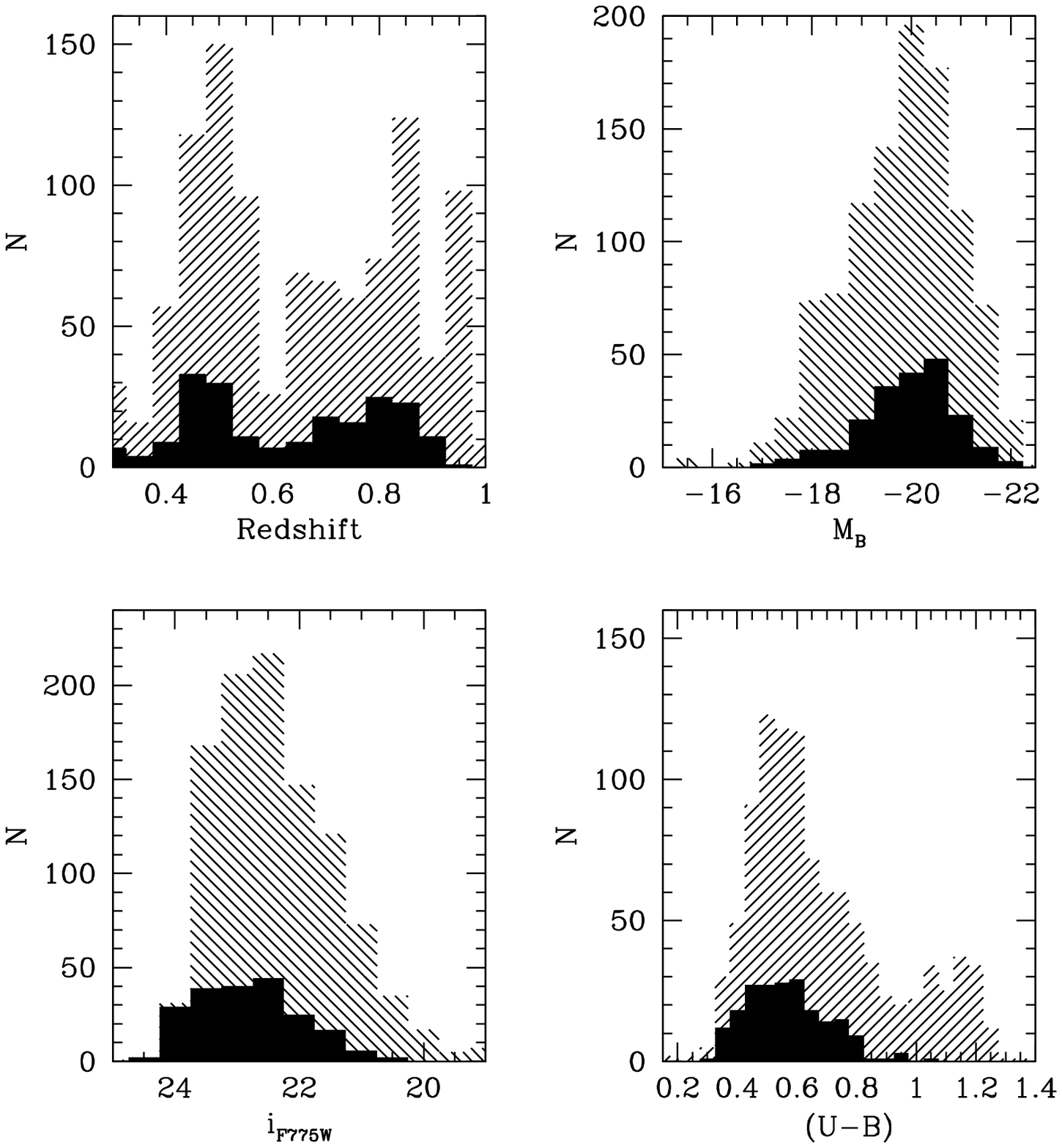}
\figcaption[hist] {Histogram of \numselect\ galaxies selected for chemical
analysis (large symbols) compared to the total 1044 objects in the
TKRS survey survey with spectroscopic redshifts $0.3<z<1.0$.  We show
the distribution as a function of redshift, observed i-band magnitude,
$M_B$, and U-B color. This figure demonstrates that galaxies selected as
suitable for chemical analysis are reasonably representative of the
larger TKRS sample in GOODS-North field in terms of their redshift
distributions, magnitudes, and luminosities.  However, the 204
selected galaxies preferentially have bluer U-B colors consistent with
higher rates of star formation, and they comprise a disproportionate fraction of galaxies
near the faint end of the survey limit at $R\sim24$.  \label{hist} }
\end{figure}

In order to assess whether the \numselect\ selected objects are
representative of the GOODS-N TKRS galaxies with spectra in the
$0.3<z<1.0$ redshift range, Figure~\ref{hist} shows histograms of
their redshifts and photometric properties.  The four panels show the
redshift distribution, $z$, the absolute B magnitudes, $M_B$, the
observed i-band (F774W) magnitudes, and the rest-frame colors, $(U-B)$.
Shaded histograms denote the selected objects while hatched histograms
show the entire TKRS sample of 1044 objects.  Examination of
Figure~\ref{hist} reveals that the \numselect\ galaxies selected for
chemical analysis are representative of the entire TKRS sample in
terms of their blue luminosities and redshift distribution.  The lower
left panel indicates that the faintest galaxies in the survey, those
near the cutoff limit, tend to be preferentially selected by our
criteria.  The lower right panel indicates that our selection criteria
also choose preferentially the bluest galaxies in the TKRS.  This
disproportionate fraction of blue galaxies is consistent with our
emission line criteria.  Those galaxies undergoing strong episodes of
star formation will have the bluest colors and will necessarily be the
ones with nebular $H\beta$, [O~II], and [O~III] emission lines.

\begin{figure}
\plotone{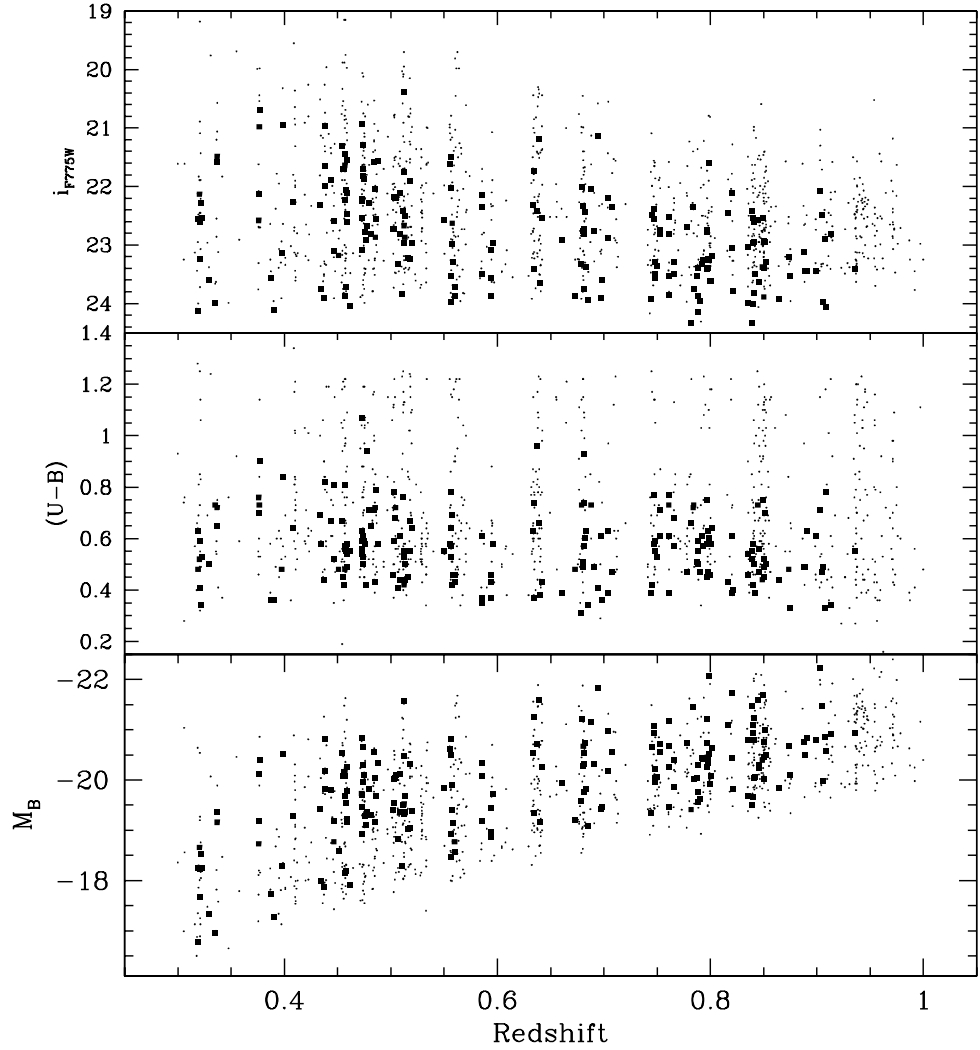}
\figcaption[scatter] 
{Distribution of apparent magnitude, U-B color, and B-band luminosity
as a function of redshift for \numselect\ galaxies selected for
chemical analysis (large symbols) and the 1044 objects in the TKRS
survey with spectroscopic redshifts $0.3<z<1.0$ (dots).
\label{scatter} }
\end{figure}

Figure~\ref{scatter} shows this selection in a slightly different way,
plotting redshift, $z$, versus photometric properties $M_B$, $(U-B)$,
and i-band magnitude.  Small dots indicate the whole set 1044 TKRS
galaxies between $0.3<z<1.0$ and solid symbols show just the
\numselect\ selected galaxies.  This figure shows that, at any given
redshift, the objects chosen for chemical analysis are representative
of the distribution of i-band magnitudes of $M_B$.   The middle panel of
Figure~\ref{scatter} shows that the selected galaxies preferentially
fall among the bluest half of the TKRS sample, for the reasons
mentioned above.  Thus, it is important to emphasize that this study
is only sensitive to the evolution of the chemical and luminous
properties of the bluest (i.e., most actively star-forming) galaxies
from $0.3<z<1.0$.

\subsection{Emission Line Measurements and Uncertainties}

We utilized the boxcar extracted (not the optimally extracted)
1-dimensional spectra made publicly available by the TKRS Team.  We
manually measured equivalent widths\footnote{The TKRS spectra are not
flux calibrated, so we use the equivalent widths of strong emission
lines in our analysis, following the prescription of Kobulnicky \&
Phillips (2003).}  of the [O~II] $\lambda\lambda3726/29$, H$\beta$,
and [O~III] $\lambda$5007 emission lines present in each of the TKRS
spectra with the IRAF SPLOT routine using Gaussian fits with variable
baseline, width and height.  The [O~II] $\lambda\lambda3726/29$
doublet, when spectrally resolved, was fit with a blend of two
Gaussian profiles.  Visual inspection of this doublet showed that the
$\lambda\lambda3726/29$ ratio, where sufficiently resolved, was always
consistent with low electron densities less than a few hundred
$cm^{-3}$.  We required that the fits to the H$\beta$ and [O~III]
$\lambda5007$ lines have the same Gaussian width.  This constraint
helped to make EW measurements more robust when one or the other of
these nebular lines was affected by night sky emission lines.
Table~\ref{src.tab} lists the measured equivalent widths and
measurement uncertainties for each line in columns 9--11. The reported
equivalent widths are corrected from the observed to the rest frame
using

\begin{equation}
EW_{rest} = EW_{observed}/(1+z).
\end{equation}

In all cases, we add 2 \AA\ to the EW of H$\beta$
as a general correction for underlying stellar absorption (see Ke03).
The EW reported for [O~III] in column 11 includes the (unmeasured)
contribution from [O~III] $\lambda$4959 using the assumption that
 $I([O~III]_{\lambda5007})/ I([O~III]_{\lambda4959})=3$ so that

\begin{equation}
EW([O~III])= 1.3\times EW([O~III]_{\lambda5007}) =  EW([O~III]_{\lambda5007}) + 
EW([O~III]_{\lambda4959}).
\end{equation}

\noindent In a few cases, the $[O~III]_{\lambda5007}$ line was hopelessly lost in the
noise from imperfectly subtracted night sky lines.  
For these objects, EW([O~III]) is measured using

\begin{equation}
EW([O~III])= 4\times EW([O~III]_{\lambda4959}),
\end{equation}

\noindent and such instances are denoted by the numeric code 2 in column 14 of 
Table~\ref{src.tab}.  For 35 low-redshift objects, equivalent widths of
the $H\alpha$ and [N~II] $\lambda$6584 lines or upper limits could also be measured.  
The ratio $EW_{H\alpha}/EW_{[N~II]\lambda6584}$ is recorded in column 
12.  Where only an upper limit on  [N~II] $\lambda$6584 could be measured,
we give the 3$\sigma$ lower limits on the $EW_{H\alpha}/EW_{[N~II]\lambda6584}$
ratio.

In 27 of the galaxies, only upper limits on [O~III] $\lambda5007$
could be measured.  These objects are located at the bottom of Table~\ref{src.tab}
and are denoted by the numeric code 3 in column 14.
To avoid a potential metallicity bias in the sample,
we include these objects in our analysis and list the 3$\sigma$ upper limits
on the EW([O~III]) in column 11 of Table~\ref{src.tab}.  The upper limits
on [O~III] (and [N~II], where possible)  are estimated by 

\begin{equation} 
EW = 3 \times RMS \times \sqrt{1.8\times npix_{FWHM}},
\end{equation}

\noindent where RMS is the rms in an offline region adjacent to the line and
$npix_{FWHM}$ is the number of pixels across the FWHM of the line profile,
typically 4-10 pixels.

Associated uncertainties on each line EW in columns 9-11
are computed taking into account both the uncertainty on the line
strength and the continuum level placement using

\begin{equation}
\sigma_{EW} = \sqrt{ {{1}\over{C^2}}\sigma_L^2 +
{{L^2}\over{C^4}}\sigma_C^2 },
\end{equation}

\noindent where $L$, $C$, $\sigma_L$, and $\sigma_C$ are the line and
continuum levels in photons and their associated $1~\sigma$ uncertainties.  We
determine $\sigma_C$ manually by fitting the baseline regions
surrounding each emission line.  We adopt
$\sigma_L=RMS\times\sqrt{1.8\times npix_{FWHM}}$.   Using this
empirical approach, the stated uncertainties implicitly include 
{\it internal} error from one emission line relative to another
due to Poisson noise, sky background, sky subtraction, readnoise, and
flatfielding.  In nearly all cases, the continuum can be fit along a
substantial baseline region, so that $\sigma_C\ll\sigma_L$.
There are, however, additional uncertainties on the {\it absolute} values
of the equivalent widths due to uncertainties on the continuum placement,
particularly for the [O~II] $\lambda$3727 line, which are difficult to 
include in the error budget. 

Using the observed H$\beta$ equivalent widths and calculated blue luminosities, 
we estimate the star formation rate for each of the TKRS galaxies in
our sample.  The H$\beta$ luminosity is estimated as

\begin{equation}
L_{H\beta}(erg/s)=5.49\times10^{31} \times 2.5^{-M_B} \times EW_{H\beta},
\end{equation}

\noindent and then the star formation rate is calculated 
following Kennicutt (1998),

\begin{equation}
SFR(M_\odot/yr)= 2.8 \times L_{H\beta} /1.26\times10^{41}
\end{equation}

\noindent The resulting star formation rate estimates, in solar masses per year,
appear in column 13 of Table~\ref{src.tab}.  This is necessarily a 
lower limit on the
star formation rate because we do not correct for extinction 
or stellar absorption.  

\subsection{Additional Galaxies }

To supplement the data on \numselect\ TKRS galaxies
analyzed here, we have collected the 64 emission line
measurements from the DEEP Groth Strip Survey (Kobulnicky
\etal\ 2003; filled symbols), and additional objects from
the Canada-France Redshift Survey from LCS03 and
Corollo \& Lilly (2002).  We use the published emission line measurements
from those original works, and include only the subset of those galaxies
where the $EW_{H\beta}$ is measured with a signal-to-noise of 8:1 or better,
in keeping with our selection criteria described above.

\subsection{Chemical Analysis \label{chem}}

Following traditional nebular diagnostic techniques (e.g., Osterbrock
1989; Pagel \etal\ 1979) and extensions of these techniques for
distant galaxies developed by Kobulnicky, Kennicutt, \& Pizagno (KKP; 1999)
and Kobulnicky \& Phillips (2003), we use the equivalent width ratios
of the collisionally excited [O~II]$\lambda$3727 and
[O~III]$\lambda$4959,5007 lines relative to the H$\beta$ Balmer series
recombination lines to estimate gas-phase oxygen abundances.  
The ratio of emission line {\it equivalent widths}, while being one step 
further removed from the
emission line {\it flux ratios} calibrated against photoionization models,
has the advantage of being reddening-independent, to first order.  

The principle diagnostic is the metallicity-sensitive emission line ratio,

\begin{equation}
\log R_{23}\equiv   \Biggr({{I_{[O~II] \lambda3727} + I_{[O~III]
\lambda4959} + I_{[O~III] \lambda 5007}}\over{I_{H\beta}}}\Biggr) \equiv x,   \label{eq_R23}
\end{equation}

\noindent formulated by Pagel \etal\ (1979) and subsequently developed and
recalibrated by many authors since (reviewed in KKP).  The $R_{23}$
ratio is sensitive to both metallicity and to the ionization state of
the gas, or ``ionization parameter''.  The ionization parameter ($q$)
is defined as

\begin{equation}
q=\frac{S_{{\rm H}^{0}}}{n}  \label{1}
\end{equation}
where $S_{{\rm H}^{0}}$ is the ionizing photon flux passing through a unit 
area, and $n$ is the local number density of hydrogen atoms. 
This ionization parameter is divided by the speed of light to give the more
commonly used ionization parameter; ${\cal U}\equiv q/c.$  
Some $R_{23}$--O/H calibrations have attempted to correct for the ionization 
parameter (e.g., McGaugh~1991;  Kewley \& Dopita~2003, hereafter KD03) by using the 
ratio of the oxygen lines, known as $O_{32}$:

\begin{equation}
\log (O_{32}) \equiv\ \log\Biggr(
	{{I_{[O~III]\lambda4959} + I_{[O~III] \lambda 5007}}
	\over{I_{[O~II]\lambda3727}}}\Biggr) \equiv y  ,  \label{eq_O32}
\end{equation}

\begin{figure}
\plotone{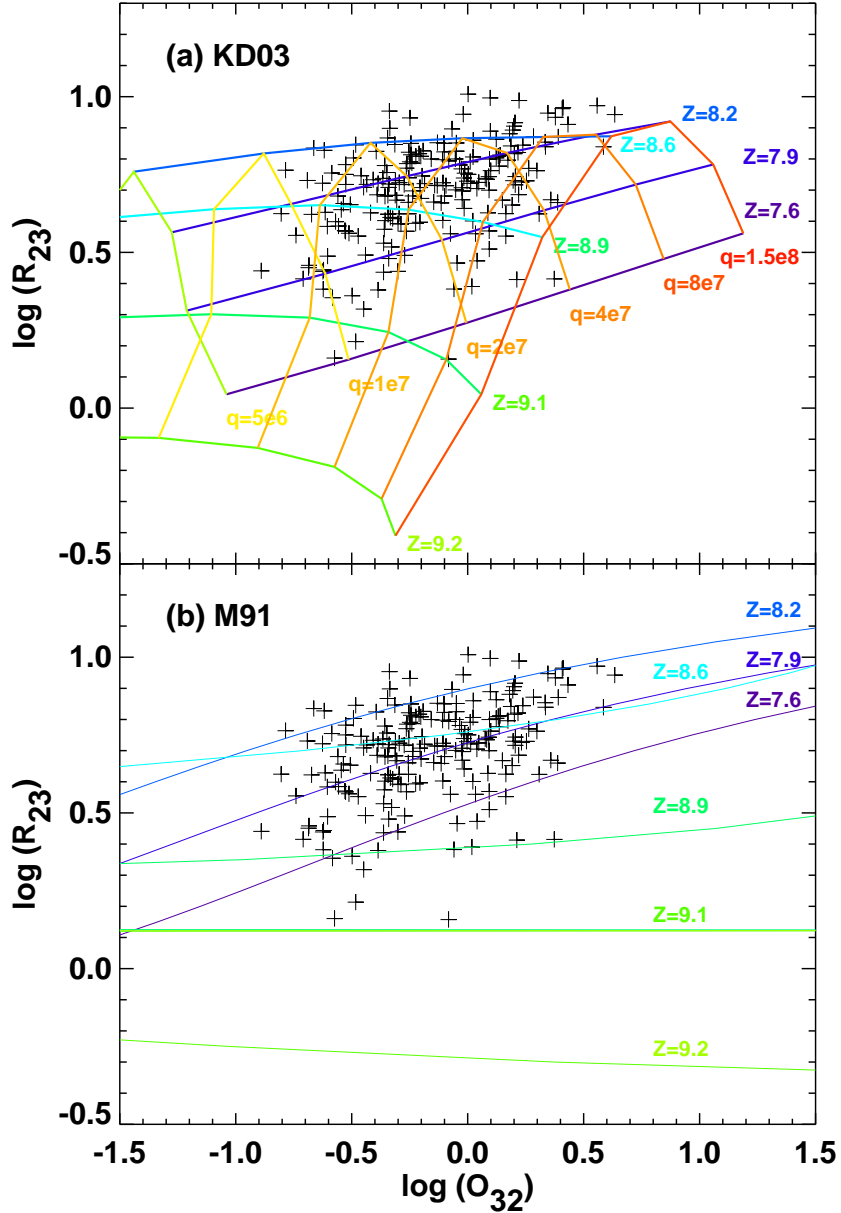}
\figcaption[R23_vs_O32] 
{The logarithm of the metallicity-sensitive line ratio $R_{23}$ 
versus the logarithm of the ionization-parameter sensitive ratio $O_{32}$ 
for the TKRS galaxies.  The colored curves represent the theoretical 
photoionization models of (a) Kewley \& Dopita~(2003) and (b) McGaugh (1991).
Models are shown for metallicities between $12+\log(\rm{O/H})=7.6$ (violet) to 
$12+\log(O/H)=9.1$ (green).  
For reference, solar metallicity is $12+\log(\rm{O/H})\sim8.7$ 
(Allende Prieto \etal\ 2001).  The Kewley \& Dopita~(2003) models 
for ionization parameters between $q=5\times10^6 - 1.5\times10^8$~cm/s 
are shown (yellow-red curves).  The TKRS data span ionization parameters
$q=1\times10^7 - 8\times10^7$~cm/s.
\label{R23_vs_O32} }
\end{figure}

In Figure~\ref{R23_vs_O32}, we show the relationship between the
$R_{23}$ ratio and $O_{32}$ for the TKRS galaxies.  The colored curves
represent the theoretical photoionization models calculated by (a)
Kewley \& Dopita (2003) and (b) McGaugh (1991).  The grids in
Figure~\ref{R23_vs_O32}a show the theoretical relationship between
$R_{23}$ and $O_{32}$ for various values of metallicity mass fraction,
$Z$, and ionization parameter, $q$.  For reference, the metal mass
fraction, $Z$, is related to the oxygen abundance, $12+\log({\rm
O/H})$, by

\begin{equation}
Z\simeq29\times 10^{[12+\log(O/H)]-12}  \label{eq_zOH}
\end{equation}

\noindent for the standard solar abundance distribution (e.g., Anders
\& Grevesse 1989) with the newer solar oxygen abundance of Allende
Prieto \etal\ (2001) which yields a solar metallicity of
$Z_\odot\simeq0.015$ and $12+\log(O/H)_\odot = 8.72$.  The grid points
for Figure~\ref{R23_vs_O32} are provided in Table~\ref{R23_O32.tab}.  
The models predict an upper
limit to $\log (R_{23})$.  This upper limit occurs because at low
metallicity the intensity of the forbidden lines scales roughly with
the chemical abundance while at high abundance the nebular cooling is
dominated by the infrared fine structure lines and the electron
temperature becomes too low to collisionally excite the optical
forbidden lines. The position of the theoretical upper limit is
similar for the Kewley \& Dopita (2003) and McGaugh (1991) models.  A
few of the TKRS galaxies have $\log(R_{23})$ maximum that is slightly
higher by $\sim0.1-0.2$~dex than the theoretical limits.  The
theoretical models were calculated assuming that the star-formation
occurred in an instantaneous burst.  This assumption may be reasonable
for \HII\ regions or for galaxy spectra dominated by one or two \HII\
regions.  However, continuous burst models may be more appropriate for
modeling the emission-line spectra of active star-forming galaxies
(e.g., Kewley \etal\ 2001).  Continuous burst models would increase the
theoretical upper limit by $\sim0.2$~dex, making the data consistent
with model limits (Kewley 2005).  We will compute new
metallicity diagnostics utilizing continuous burst models in a future
paper.

\begin{figure}
\plotone{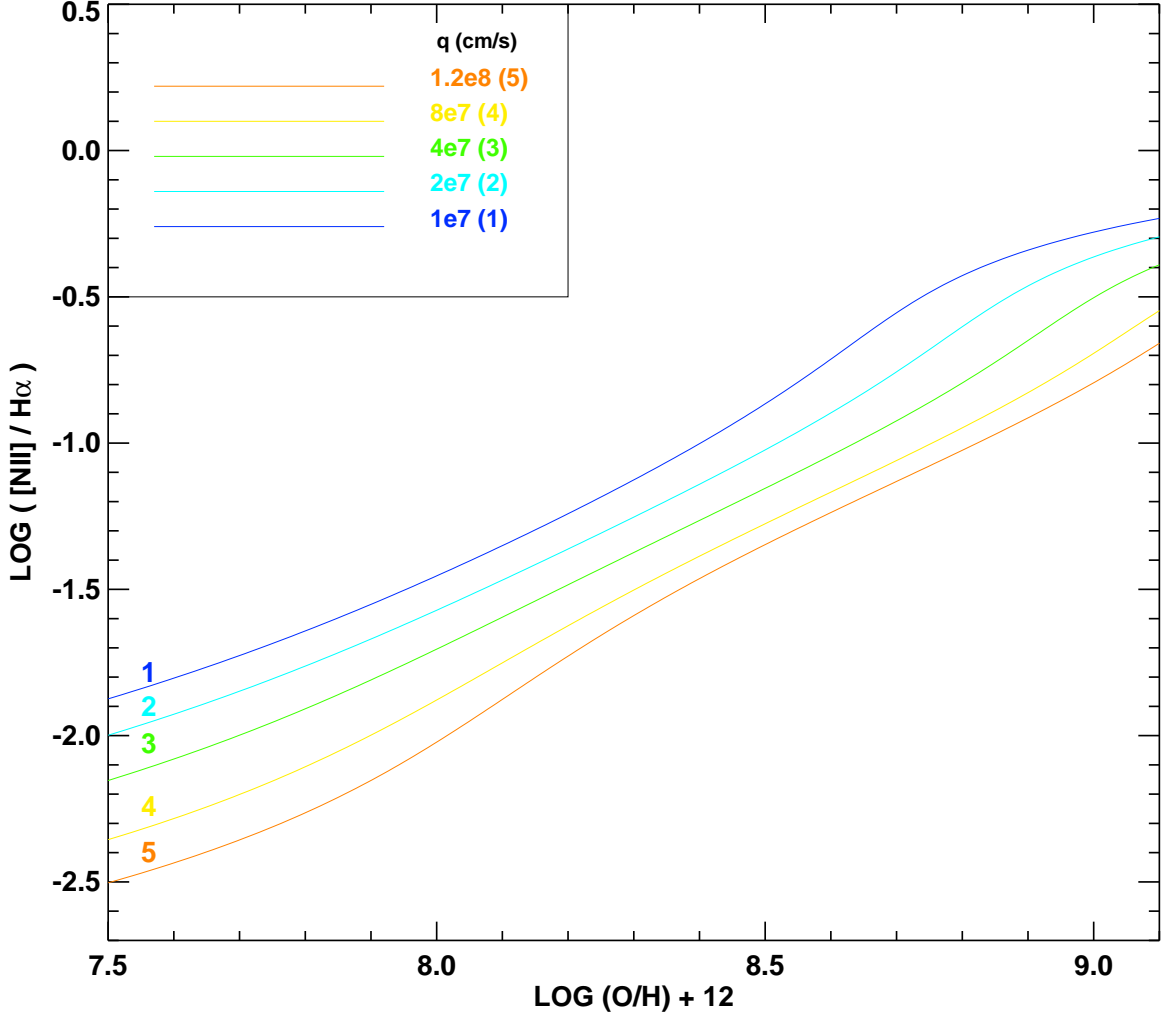}
\figcaption[NIIHa_fit] 
{The metallicity-sensitive line ratio \NIIHa\ versus the metallicity 
$12+\log(O/H)$.  The colored curves show our new parameterization (equation~\ref{eq_NIIHa}) to the 
theoretical photoionization models of Kewley \& Dopita~(2003) 
 for various values of ionization parameter, $q$, in cm s$^{-1}$ 
shown in the legend. Note that the \NIIHa\ ratio is
very sensitive to ionization parameter and is a not a robust
indicator of metallicity.  The metallicity corresponding
to any particular value of \NIIHa\ spans 0.4---0.6 dex in O/H, depending on
ionization parameter.  
\label{NIIHa_fit} }
\end{figure}

\begin{figure}
\plotone{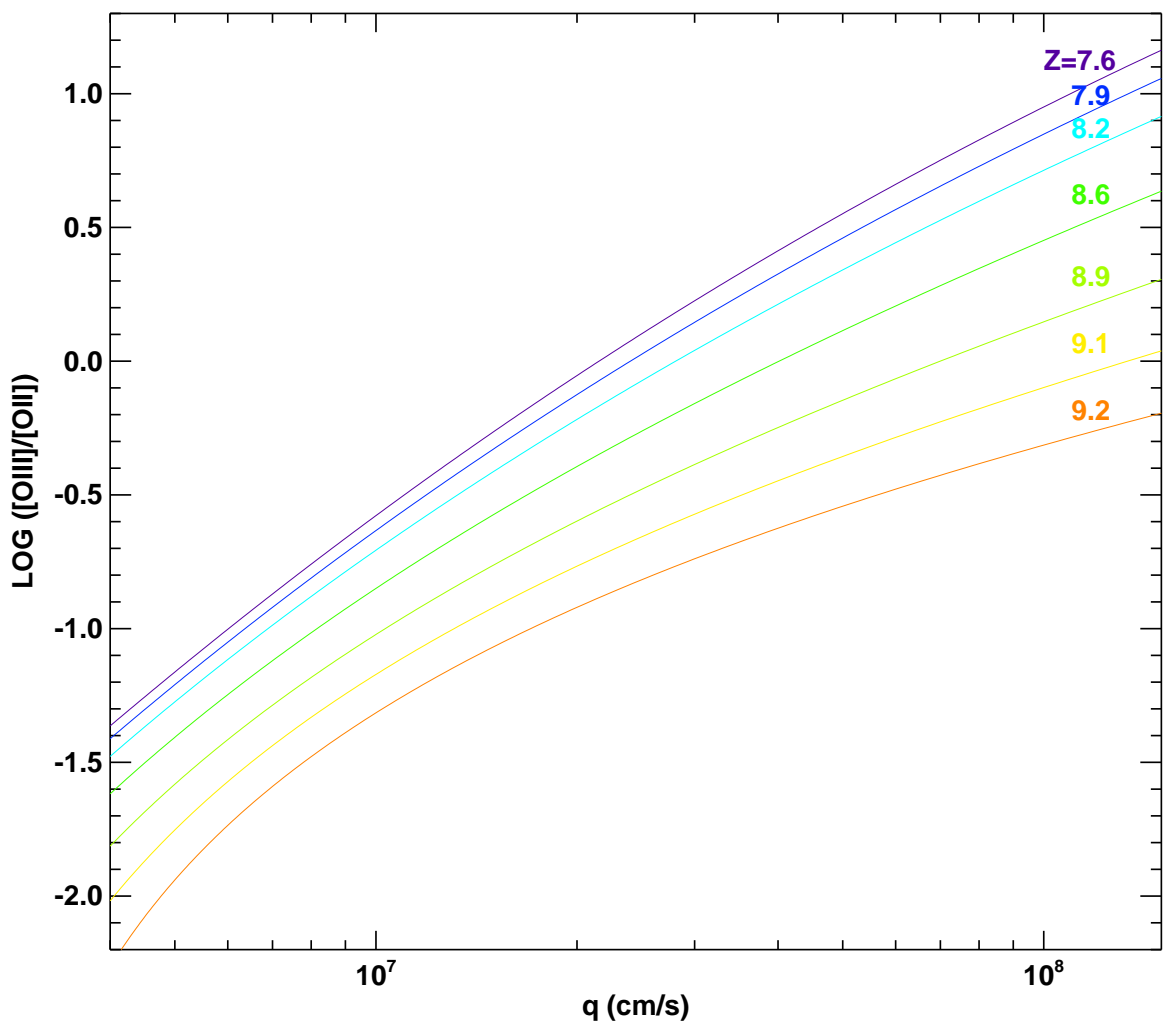}
\figcaption[OIIIOII_fit] 
{The  \OIIIOII\ ratio versus ionization parameter, $q$, in cm s$^{-1}$.  
The colored curves show our new parameterization (equation~\ref{eq_OIIIOII}) 
to the  theoretical photoionization models of Kewley \& Dopita~(2003) 
 for various values of metal mass fraction, $Z$, shown.
The relation between $Z$ and O/H is discussed in Section~\ref{chem} and
Equation~\ref{eq_zOH}. 
\label{OIIIOII_fit} }
\end{figure}

Figure~\ref{R23_vs_O32} illustrates the difficulty in using \R23\ to
diagnose metallicity.  Not only is \R23\ sensitive to ionization
parameter but it is double valued in terms of the metallicity.  To
break the degeneracy, other metallicity-sensitive line ratios are
required.  For the 30 galaxies in our sample with measured \NIIHa\
ratios, we calculate an initial metallicity using the
\NIIHa$--$metallicity formulation from KD03.  A further 2 galaxies have small \NIIHa\ upper limits
that allowed us to break the \R23\ degeneracy.  For the potential
ionization parameter and metallicity range of our sample
(Figure~\ref{R23_vs_O32}), the KD03 \NIIHa-metallicity relation can be
parameterized as

\begin{eqnarray}
12+\log({\rm O/H}) & =  & 7.04 + 5.28 X_{NII}+6.28 X_{NII}^2+2.37 X_{NII}^3\nonumber\\
 & & -\log(q)(-2.44-2.01X_{NII}  -0.325 X_{NII}^2+0.128 X_{NII}^3)\nonumber\\
 & & +10^{X_{NII}-0.2}\log(q)(-3.16+4.65X_{NII})  \label{eq_NIIHa}
\end{eqnarray}

\noindent where $X_{NII}=\log EW([N~II])_{6584}/EW(H\alpha)$.   
The ionization parameter is calculated from the KD03 \OIIIOII--$q$ relation 
that we parameterize as

\begin{equation}
\log(q)=\frac{32.81 -1.153 y^2 + [12+log(O/H)](-3.396 -0.025 y +  0.1444 y^2)}{
        4.603 -0.3119 y -0.163 y^2 + [12+log(O/H)](-0.48 + 0.0271 y+ 0.02037 y^2)} \label{eq_OIIIOII}
\end{equation}

\noindent where $y=\log(O_{32})$.  Figures~\ref{NIIHa_fit} and \ref{OIIIOII_fit} 
show the parameterizations in equations~\ref{eq_NIIHa} and \ref{eq_OIIIOII} 
for various values of ionization parameter.
Equation~\ref{eq_NIIHa} is only valid for
ionization parameters $q$ between $5\times10^6 - 1.5\times10^8$~cm/s.
KD03 show that for log(\NIIHa)$>-0.8$, the \NIIHa$--$metallicity
relationship breaks down.  Although such high \NIIHa\ ratios indicate
that the metallicity is on the upper branch ($12+\log({\rm
O/H})>8.4$), \NIIHa\ cannot be used to estimate a metallicity in this
regime.  At lower \NIIHa\ values, the \NIIHa\ ratio is less sensitive
to metallicity and is more dependent on the ionization parameter than
the \R23\ ratio.  Therefore, \NIIHa\ should only be used as a crude
initial estimate for a more sensitive metallicity diagnostic such as
\R23.

Because the \NIIHa\ ratio depends strongly on the ionization parameter
and the \OIIIOII\ ratio depends on metallicity, we iterated
equations~\ref{eq_NIIHa} and \ref{eq_OIIIOII} until the metallicity
varied less than the model errors ($\sim0.1$~dex).  Typically 2-3
iterations were required.  The resulting metallicity estimates
indicate that 29/32 (91\%) of the galaxies with useable \NIIHa\ ratios
or upper limits lie on the upper \R23\ branch ($12+log(O/H)>8.4$).  In
column 20 of Table~\ref{src.tab} we list the oxygen abundance derived
from the strength of the [N~II] $\lambda$6584/H$\alpha$ equivalent
width ratios for the 35 (necessarily the lowest redshift) galaxies
where these lines can be measured.  In many cases, only 3$\sigma$
upper limits on the [N~II] $\lambda$6584 equivalent widths are
measured, so only upper limits on the oxygen abundances are given.  Of
the galaxies where [N~II] can be measured, only 2 of the least
luminous galaxies ($M_B\sim-17$) have [N~II] $\lambda$6584/H$\alpha$
ratios consistent with very low metallicities that would place them on
the lower branch of the $R_{23}$--O/H calibration.  These objects are
noted in Table~\ref{src.tab}.

For the TKRS galaxies without \NIIHa\ ratios, we are unable to break
the \R23\ degeneracy.  We make the motivated assumption (see
Kobulnicky \& Zaritsky 1999, Kobulnicky \etal\ 2003) that these
galaxies fall on the upper-branch ($12+log(O/H)>8.4$) of the
double-valued $R_{23}-O/H$ relation, consistent with the majority of
galaxies in our sample with \NIIHa\ ratios.  We next compute the
oxygen abundances of TKRS galaxies using several different,
independent but related \R23--O/H calibrations from the literature.

\subsubsection{Zaritsky, Kennicutt, \& Huchra 1994, (ZKH)}

Zaritsky, Kennicutt, \& Huchra 1994, (ZKH)
provide a simple analytic relation between O/H and
$R_{23}$ only without regard to ionization parameter:

\begin{equation}
12+\log({\rm O/H})_{\rm ZKH}  =  9.265-0.33\,{R}_{23}-
                       0.202\,{R}_{23}^{2}-
                       0.207\,{R}_{23}^{3}-
                       0.333\,{R}_{23}^{4}   
\end{equation}

\begin{figure}
\plotone{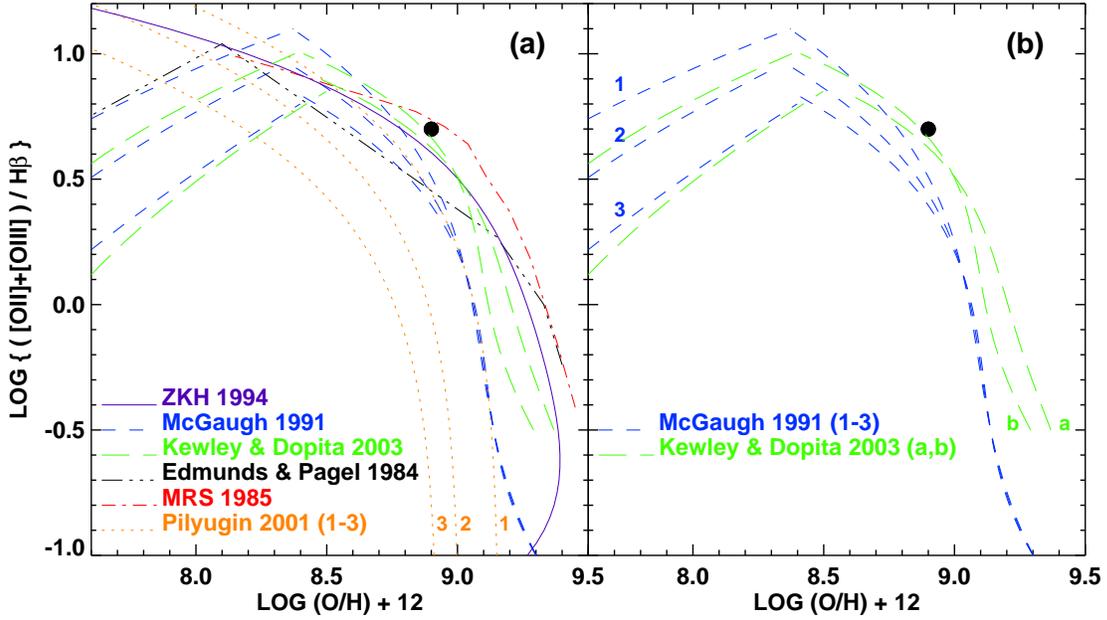}
\figcaption[R23OH] 
{Metallicity calibration between strong line ratio $R_{23}$ and oxygen
abundance, $12+\log(O/H)$ from several different authors as indicated in the
legend.   A solid circle marks the Orion Nebula value (based on data of
Walter, Dufour, \& Hester 1992).  For reference, the currently accepted 
value of solar abundance is $12+\log(O/H)_\odot=8.7$ 
(Allende Prieto, Lambert, \& Asplund 2001).  Panel (a) shows 
the variation in \R23\ calibrations from the literature, while panel (b) shows 
the M91 and KD03 calibrations. 
The  three curves labeled 1,2,3 denote the Pilyugin 2001 and M91
models corresponding to $\log({\rm O_{32}})=1,0,-1$ respectively.  Two curves
labeled a,b denote the Kewley \& Dopita (2003) models corresponding to ionization
parameters of $q=1\times10^{7}$ and $q=1.5\times10^{8}$~cm/s respectively.
\label{R23OH} }
\end{figure}

\noindent The ZKH formula is an average of three previous calibrations 
in the literature, namely Dopita \& Evans (1986), 
McCall, Rybski \& Shields (1985), and Edmunds \& Pagel (1984).
Figure~\ref{R23OH} shows these relations.  The ZKH
average relation was compared with a sample of disk \HII\ regions
with metallicities \OH $> 8.35$.  As a result, the ZKH calibration
is only suitable for \HII\ regions in the metal-rich regime.

\subsubsection{McGaugh 1991 (M91)}

McGaugh (1991) calibrated the relationship between the $R_{23}$ ratio
and gas-phase oxygen abundance using \HII\ region models from the
photoionization code CLOUDY (Ferland \& Truran 1981).  McGaugh's
models include the effects of dust and variations in ionization
parameter.  Adopting the analytical expressions of McGaugh (1991, 1998
as expressed in KKP) which are based on fits to photoionization models
for the metal-rich (upper) branch of the $R_{23}$--O/H relation. In
terms of the reddening corrected line intensities, this relation is

\begin{eqnarray}
12+log(O/H)_{\rm M91,upper} & = & 12 -2.939-0.2x-0.237x^2-0.305x^3-0.0283x^4-  \nonumber\\
& & 	y(0.0047-0.0221x-0.102x^2-0.0817x^3-0.00717x^4).
\end{eqnarray}

\noindent Figure~\ref{R23OH} shows graphically the relation between
$R_{23}$ and O/H for the McGaugh (1991) and other calibrations from
the literature.  A circle marks the Orion Nebula value (based on data of
Walter, Dufour, \& Hester 1992) which is in excellent agreement with
the most recent solar oxygen abundance measurement of
$12+log(O/H)_\odot=8.7$ (Prieto, Lambert, \& Asplund 2001).  Oxygen
abundances computed in this manner appear in Table~\ref{src.tab}
column 17, along with a $1\sigma$ uncertainty computed by propagating
the uncertainties on the emission line equivalent widths.  This uncertainty
estimate does not include the error introduced by the model uncertainties 
in the theoretical calibrations (typically $\sim0.1$~dex).

\subsubsection{Pilyugin (2001)}

Pilyugin (2001) developed an $R_{23}$--O/H calibration based on a
sample of \HII\ regions with measurements of the \OIII~$\lambda4363$
auroral line.  The \OIII~$\lambda4363$ auroral line provides a
``direct'' measurement of the electron temperature of the gas, and
therefore, the metallicity.  The use of the \OIII~$\lambda4363$
auroral line to derive metallicities is known as the T$_{e}$ method.
Pilyugin (2001) calculated direct metallicities for a sample of \HII\
regions spanning $8.2\lesssim12+\log(O/H)\lesssim8.6$.  His resulting
$R_{23}$--O/H calibration provides three curves, depending on a
parameter $P$ that accounts for the range in ionization parameter in
the \HII\ regions.  Pilyugin was unable to provide a fit to \HII\
regions for $12+\log(O/H)\gtrsim8.6$ because high metallicity galaxies
have weak or undetectable
\OIII~$\lambda4363$ (see Stasinska 2002 for a discussion).  Therefore,
Pilyugin extrapolated the $8.2\lesssim12+\log(O/H)\lesssim8.6$ curves
to higher metallicities. The Pilyugin curves are discussed further in
Kennicutt, Bresolin, \& Garnett (2003).

\subsubsection{Kewley \& Dopita (2003)}

Kewley \& Dopita (2003) provide a suite of abundance calibrations
depending on the availability of particular nebular emission-lines.  Their
calibrations are based on a combination of stellar population
synthesis models (P\'{e}gase and STARBURST99) and detailed
photoionization models using the MAPPINGS code (Sutherland et
al. 1993).  Like M91, the KD03 models include the effects of dust.
KD03 include separate calibrations for ionization parameter.  KD03
point out that the $R_{23}$ curve is dependent on ionization
parameter, while the common ionization parameter diagnostic
($O_{32}$) depends on metallicity.  They advocate the use of an
iterative scheme to solve for both quantities if only [O~III], [O~II]
and H$\beta$ are available.  We provide a new parameterization of the
KD03 \R23\ method (from Section 4.3 in in KD03) with a similar form to
the M91 calibration to facilitate metallicity estimation and
comparisons to estimates made using M91.  For the potential ionization
parameter and metallicity range of our sample
(Figure~\ref{R23_vs_O32}), the lower branch ($12+\log(O/H)<8.4$) is
parameterized by:

\begin{equation}
12+\log({\rm O/H})_{\rm KD03,lower} = 9.40 + 4.65 x -3.17 x^2 - \log(q) (0.272+ 0.547 x -0.513x^2)
\end{equation}

\noindent where $x=\log(R_{23})$.  The upper branch (12+log(O/H)$\ge8.4$) is parameterized by:

\begin{eqnarray}
12+\log({\rm O/H})_{\rm KD03,upper} & = & 9.72  - 0.777 x  -0.951 x^2 - 0.072 x^3
               -0.811 x^4 - \nonumber\\
               & & \log(q) (0.0737 -0.0713 x  -0.141 x^2+
               0.0373 x^3 -0.058 x^4). \label{eq_R23_KD03}
\end{eqnarray}

\begin{figure}
\plotone{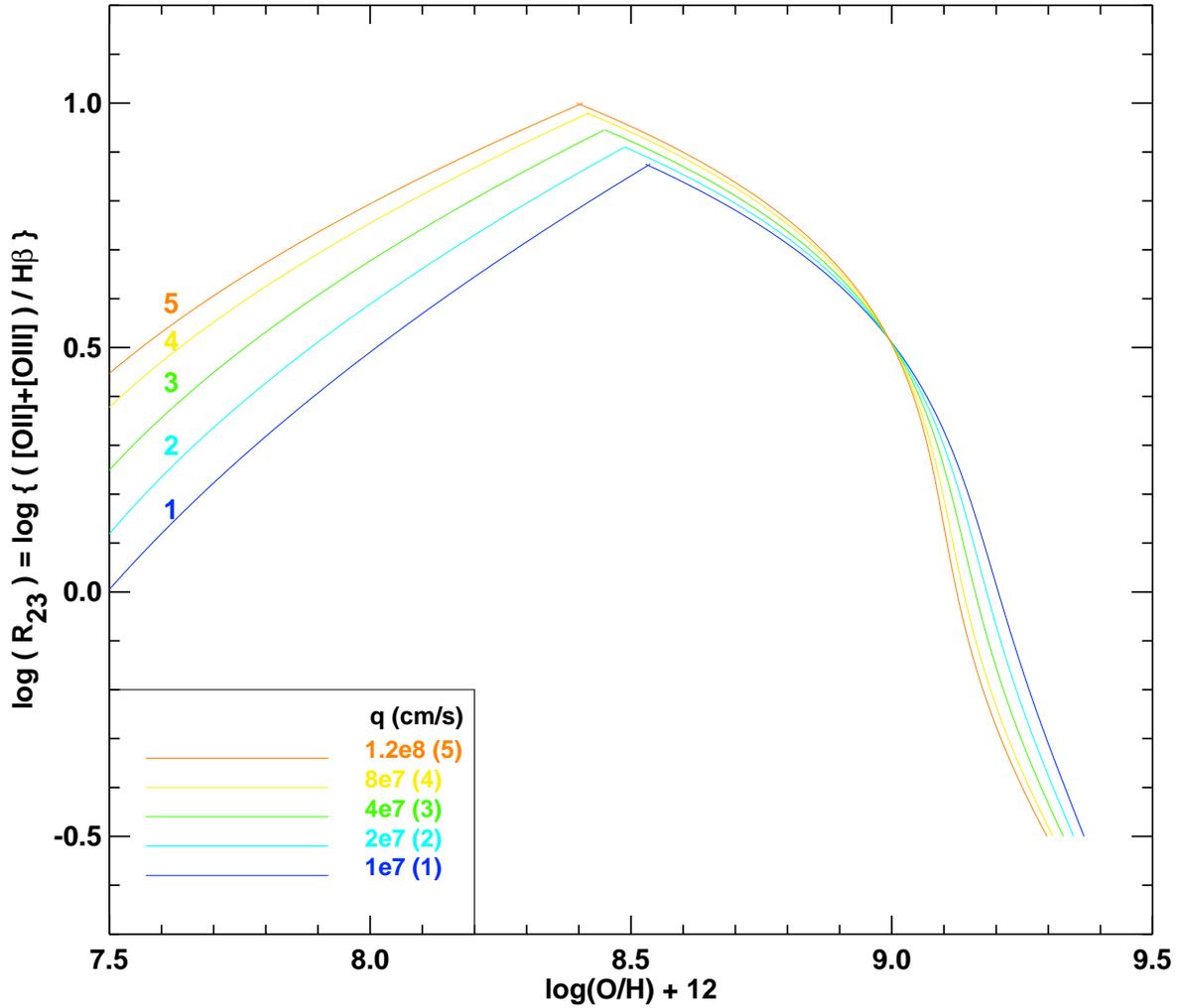}
\figcaption[R23_fit] 
{The metallicity-sensitive line ratio \R23\ versus the metallicity 
$12+\log(O/H)$.  The colored curves show our new parameterization (equation~\ref{eq_R23_KD03}) to the 
theoretical photoionization models of Kewley \& Dopita~(2003) 
 for various values of ionization parameter, $q$, in cm~s$^{-1}$  
shown in the legend. 
\label{R23_fit} }
\end{figure}

Figure~\ref{R23_fit} shows graphically the relation between the
metallicity and \R23\ from equation~\ref {eq_R23_KD03} for various
values of ionization parameter $q$ .  The ionization parameter, $q$,
is found using equation~\ref{eq_OIIIOII} in an iterative manner.
Typically 2-3 iterations were required to reach convergence.  This new
parameterization is an improvement over the tabulated model
coefficients of the KD03 calibration because (1) the new calibration
does not fix the ionization parameter or metallicity to the finite set
of model values during iteration, and (2) there is an increased
sensitivity to the metallicity around the local maximum ($Z=8.4$)
because we have introduced different equations for the two branches
rather than being limited to one equation. Oxygen abundances
calculated via this method are given in column 18 of
Table~\ref{src.tab}.  This parameterization should be regarded as an
improved, implementation-friendly approach to be preferred over the
tabulated $R_{23}$ coefficients of KD03.

\subsubsection{``Best'' Adopted Oxygen Abundances}

\begin{figure}
\plotone{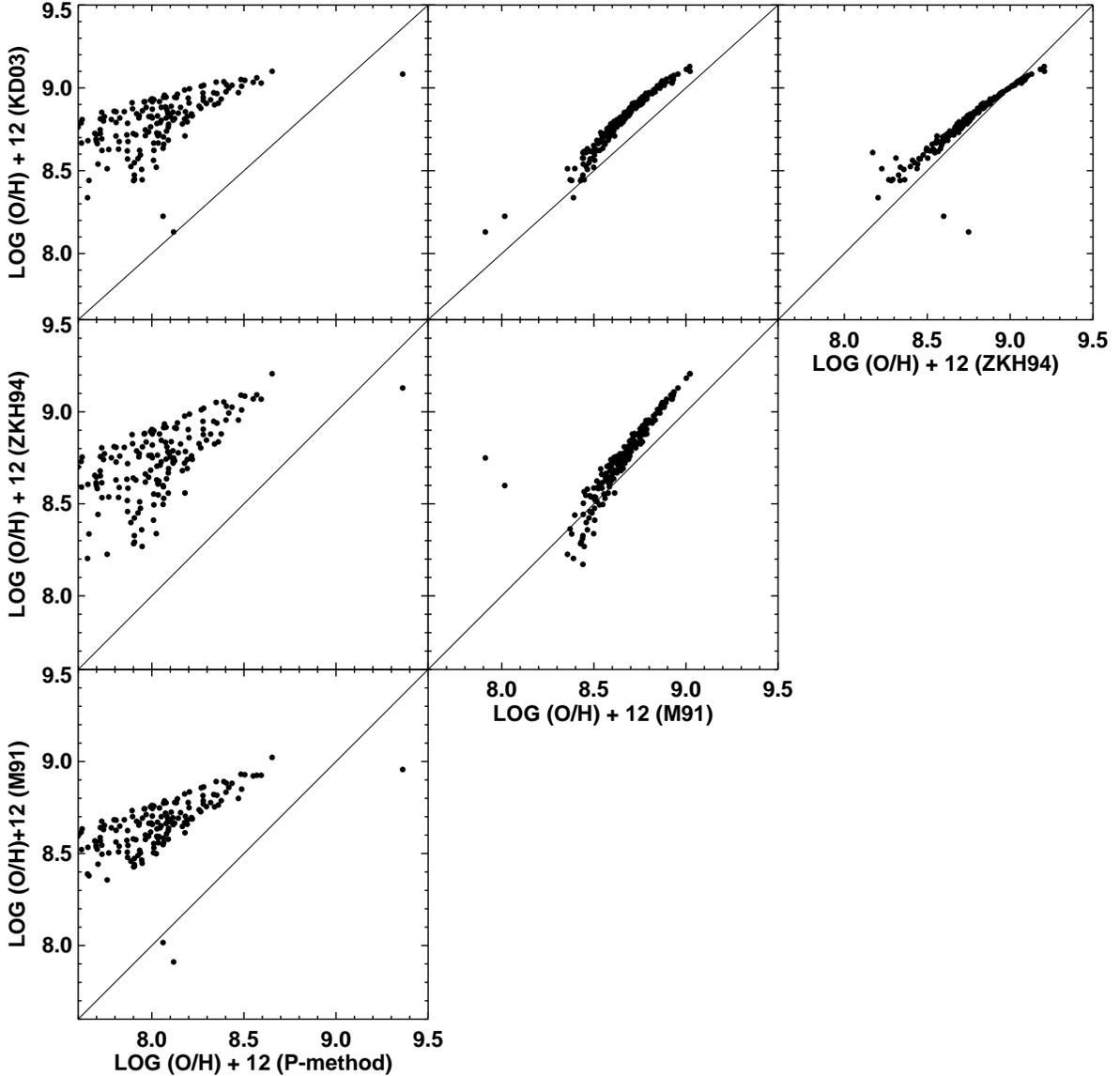}
\figcaption[compare1] 
{Comparison of difference oxygen abundance computation methods for
TKRS galaxies.  The solid y=x line indicates where the data would lie
if the different metallicity estimates reach agreement.  The
`P-method' proposed by Pilyugin (2001) produces a strong systematic
offset and significant scatter compared to the other three
calibrations: McGaugh (1991; M91), Zaritsky, Kennicutt \& Huchra
(1994; ZKH94), and Kewley \& Dopita (2003; KD03).  Overall, the M91,
ZKH94 and KD03 methods (with our new parameterization of the KD03
\R23\ and \OIIIOII models) produce metallicity estimates that reach
agreement to within the error estimates for the calibrations ($\sim
0.1$~dex each).  The ZKH94 method should not be used for
$12+\log(O/H)<8.35$ because the ZKH94 method places these galaxies on
the upper \R23\ branch.
\label{compare1} }
\end{figure}

Of all the above methods for oxygen abundance computation, many
arguments could be made for which are the ``best'' or most
``accurate''.  In Figure~\ref{compare1} we compare the `P-method',
M91, ZKH and KD03
\R23\ metallicity estimates for the TKRS data.  
Differences between published calibrations shown in
Figure~\ref{compare1} serve to illustrate the magnitude and severity
of the possible systematic errors introduced by the different
calibrations.  The uncertainties from any published calibration method
are dominated by systematic uncertainties and/or biases in the data
and/or models used to construct the calibration.  
The `P-method' produces a strong systematic offset and large scatter
in the metallicity estimates compared to the three other calibrations.
This offset probably occurs because the P-method was not calibrated
using any data or theoretical models for the metallicity range of our
sample.  We therefore do not use the P-method in our preferred
metallicity estimates.

The three remaining calibrations show smaller systematic offsets that
are consistent with the error estimates of the calibrations
($\sim0.1$~dex).  Principle differences among the models include
different photoionizing radiation fields from the various stellar
atmospheres, stellar libraries, and stellar tracks.   Different
photoionization models employ various atomic data and dust
prescriptions.  Different calibration data from observations of
Galactic or extragalactic \HII\ regions over a range of metallicities
and ionization parameters may also affect the resulting calibrations.
Analyzing the nuances and resolving the differences between the
published strong-line---metallicity calibrations is beyond the scope
of this work.  The exact choice of metallicity calibration is not
crucial to the first of the two goals in this paper.  {\it Relative}
differences in metallicity between samples at different redshifts will
not be sensitive to the exact form of the $R_{23}$--O/H calibration
adopted.  Section~\ref{LZ.sec} below deals primarily with this
question.  Any of the three above methods which require only
measurements of $R_{23}$ and $O_{32}$ will suffice for discerning
trends with redshift.

Kennicutt, Bresolin, \& Garnett (2003) and Garnett, Kennicutt, \&
Bresolin (2004) present evidence from observations of metal-rich \HII\
regions in M~51 and M~101 that there is a discrepancy
of several tenths of a dex (factor of 2 or more) between
the metallicities derived using traditional strong line methods and
models compared to methods using 
direct measurements of [N~II] electron temperatures.
See these works for an extensive discussion regarding
possible systematic effects in the $R_{23}$--O/H calibrations.  {\it
hence, absolute metal abundances may have systematic uncertainties of
0.2--0.5 dex, particularly at the high metallicity end of our sample.}
Until these initial results are confirmed and a more robust
calibration is available, we proceed to adopt a combination of the
KD03 and M91 strong line formulation.
 
For the purposes of providing the most meaningful metallicities to aid
in modeling of distant galaxies, we adopt a final ``best estimate''
oxygen abundance by averaging the KD03 and M91
\R23\ methods; 
$12+\log({\rm O/H})_{\rm avg}= 1/2  [12+\log({\rm O/H})_{\rm KD03}+\log({\rm O/H})_{\rm M91}]$.  
We use the new KD03 parameterization (equations~\ref{eq_R23_KD03} and
\ref{eq_OIIIOII}).  Because the ZKH parameterization is based upon an
ad-hoc average of 3 relatively old models and calibrations, we do not
consider it for our best estimate in light of improvements to
photoionization models (see for example Dopita \etal~2000, Kewley et
al.~2001 and references therein).

The resulting average abundances appear in Table~\ref{src.tab}.   
For $12+\log({\rm O/H})>8.4$, the average of the KD03 and M91 methods 
can be approximated by the following simple form:

\begin{eqnarray}
12+\log({\rm O/H})_{\rm avg-upper}& \sim & 9.11 -0.218 x -0.0587 x ^2
-0.330 x^3 -0.199 x ^4 \nonumber\\ & & -
y(0.00235-0.01105x-0.051x^2-0.04085x^3-0.003585x^4) \label{eq_best}
\end{eqnarray}

\noindent where $x=\log(R_{23})$ (equation~\ref{eq_R23}) and $y=\log(O_{32})$
(equation~\ref{eq_O32}).  Note that this equation is only valid for
the upper branch, and for the range of \R23\ and \O32\ values covered by our
sample (Figure~\ref{R23_vs_O32}).  

\subsection{Comparison with other properties}

Figure~\ref{multi} shows the relations between the ``best estimate"
oxygen abundance, $H\beta$ equivalent width, star formation rate,
absolute B-band magnitude, U-B color, and redshift for the \numselect\
selected TKRS galaxies. This Figure illustrates that the colors and
emission line equivalent widths of selected TKRS galaxies are
distributed approximately uniformly across the full observed range of
parameter space within each redshift bin.  It also shows that galaxies
at increasingly higher redshifts tend to be more luminous and have
higher star formation rates.\footnote{Note that the absolute blue
magnitudes and star formation rates are closely coupled parameters
since $M_B$ is used, along with $EW_{H\beta}$, to estimate the $SFR$.
They are not independent parameters. } This is a type of selection
effect due to the magnitude limited nature of the spectroscopic TKRS
sample.

\section{The L-Z Relation at $0.3<z<1.0$. \label{LZ.sec}}

\subsection{Evolution of the L-Z Relation}

The addition of 204 new metallicities at $0.3<z<1.0$ allows us to
explore more robustly than before the chemo-luminous evolution of
galaxies over the last 8 Gyr. Figure~\ref{LZ1} shows the oxygen
abundance, $12+\log(O/H)_{M91}$ versus $M_B$ for four redshift ranges.
Filled symbols show the DGSS data from Kobulnicky \etal\ (2003),
crosses denote the CFRS data of LCS03 and CL01, and open symbols
denote the TKRS data.  Stars in the lower right panel are the $z>2$
Lyman break galaxies from Kobulnicky \& Koo (2000) and Pettini \etal\
(2001).  The dashed lines, which are the same in each panel, represent
the fits to local emission line galaxy samples as described in
Kobulnicky \etal\ (2003).  The solid lines are fits to the DGSS data
alone.  The new TKRS data are in good agreement with the previous L-Z
relations found for each redshift interval.  It becomes possible, for
the first time with these data, to see the L-Z correlation at
redshifts $z>0.8$ in the lower right panel.  The most significant
departure from the DGSS L-Z correlation seen in the TKRS data occurs
in the $0.4<z<0.6$ bin (lower left panel).  There are galaxies
located to the faint/metal-rich side of the best fit line in the TKRS
sample which do not appear in the DGSS sample.  This scatter is
present to a lesser extent in the $0.6<z<0.8$ range (upper right
panel).  Examination of these objects in Table~\ref{src.tab} reveals
these to be mostly objects with extreme values of the ionization
parameter indicator, $O_{32}$.  Most of these galaxies have $\log
O_{32}<-0.5$, placing them in an regime where the correlation between
$R_{23}$ and $EWR_{23}$ (KP03, Figure 5) has a large dispersion and a
systematic deviation from unity, making these points additionally
uncertain.  The conclusion of Figure~\ref{LZ1} is to highlight the
overall good correspondence of TKRS metallicities with other surveys
for similar magnitude and redshift ranges.

Figure~\ref{LZ2} shows our adopted oxygen abundances,
$12+\log(O/H)_{avg}$, for the DGSS+CFRS+TKRS galaxies versus $M_B$ in
each of four redshift intervals. Lines now represent fits to the
combined data in each redshift interval.  The dotted line is a fit of
$M_B$ on O/H, the dashed line is the inverse fit of O/H on
$M_B$. Since neither metallicity uncertainties nor the magnitude
uncertainties are well characterized, we also use the method of linear
bisectors described by Isobe \etal\ (1990), shown by the solid line.
Parameters for the linear fits appear in Table~\ref{fit.tab}.  The
fits show that, regardless of the fitting method adopted, there is
evidence that the slope of the L-Z correlation changes with redshift.
For the linear bisector fits, the L-Z relation evolves monotonically
from $12+\log(O/H)=-0.164\pm0.031\times M_B + 5.57$ in the lowest
redshift bin to $12+\log(O/H)=-0.241\pm0.030\times M_B + 3.73$ in the
highest redshift bin.  For the O/H on $M_B$ fits, the L-Z relation
evolves monotonically from $12+\log(O/H)=-0.074\pm0.032\times M_B +
7.20$ in the lowest redshift bin to $12+\log(O/H)=-0.134\pm0.032\times
M_B + 5.97$ in the highest redshift bin.  This change is driven mainly
by the appearance of a population of relatively luminous
($M_B\sim-20$) but metal-poor ($12+log(O/H)$=8.5--8.6) galaxies in the
two highest redshift bins. Ke03 reported this change in slope of the
L-Z relation using data out to $z=0.8$, and Figure~\ref{LZ2} shows
that this trend continues to higher redshifts.  Based on plausible
galaxy evolution models, optical morphologies, and their present
metallicities, this population of galaxies is likely to evolve at
roughly constant luminosity into comparatively metal-rich disk
galaxies in the local universe rather than fade into ``dwarf''
galaxies (see Ke03, LCS03).

The evolution of mean galaxy metallicity with redshift can be seen
more easily in Figure~\ref{zOH} which shows oxygen abundance,
$12+\log(O/H)_{avg}$, versus redshift for three different luminosity
bins.  The upper row displays only distant galaxies, $z>0.3$.  Lines
show least squares fits to the data with errors in O/H only.  The left
panel shows the lowest luminosity galaxies with $-18.5>M_B>-19.5$.
The slope of the least squares linear fit is -0.19$\pm0.10$ dex per
unit redshift.  The linear correlation coefficient for the 60 galaxies
in this panel is -0.201, indicating that the probability of obtaining
such a strong correlation at random is 12\%.  For the middle panel
showing 100 galaxies in the luminosity range $-19.5>M_B>-20.5$, the
best fit slope is -0.19$\pm0.08$ dex per unit redshift.  The
correlation coefficient is -0.228, indicating that the probability of
exceeding this degree of correlation by chance is 2\%. For the right
panel showing 69 galaxies in the luminosity range $-20.5>M_B>-21.5$,
the best fit slope is -0.18$\pm0.09$ dex per unit redshift.  The
correlation coefficient is -0.197, indicating that the probability of
exceeding this degree of correlation by chance is 10\%.  Note that the
relation between redshift and mean galaxy metallicity may not be
(indeed, is theoretically not expected to be) a linear one if the star
formation rate is not a linear function of redshift (e.g., Somerville
2001).  However, the observed dispersion and limited range of
luminosity and redshift of the current data do not warrant additional
parameters in the fit.

The lower row of Figure~\ref{zOH} shows the same galaxies as the upper
row, but with the addition of local $z=0$ galaxies from Kennicutt
(1992) and Jansen \etal\ (2001) used by Ke03 to define the local L-Z
relation.  Lines show least squares fits to the sample.  The slopes
are now somewhat smaller, 0.14$\pm0.05$ dex/z for the low- and
intermediate-luminosity bins and 0.13$\pm0.05$ dex/z in the
high-luminosity bin.  This high-luminosity bin lacks the population of
low-redshift ($0.3<z<0.6$), low-metallicity (12+log(O/H)<8.7) galaxies
present in the middle and left panels.  It is this population, or
rather the lack thereof, which is responsible for the change in slope
of the L-Z relation in Figures~\ref{LZ1} and \ref{LZ2}.  The
metallicities of the most luminous objects could be described as
reaching a plateau near $12+log(O/H)=8.9$ with increasing scatter to
lower oxygen abundances at larger redshifts.  Such a plateau is
expected in chemical evolution models as galaxies expend their gas
supplies near the cessation of star formation activity.  However, this
plateau may also be a consequence of the $R_{23}$ to O/H formulation
(Figure~\ref{R23OH}) which asymptotes at high metallicities, folded
with observational selection effects which will preferentially exclude
very high metallicity objects which have very weak [O~III] emission
lines.

The mean rate of metal enrichment observed in Figure~\ref{zOH} is at
least 0.14 dex per unit redshift for all luminosity bins.  This rate
of metal enrichment is significantly greater (2-3$\sigma$) than the
increase of $0.08\pm0.02$ dex estimated by Lilly, Carollo, \& Stockton
(2003) over the same redshift interval.  The smaller
${d(O/H)}\over{dz}$ measured by LCS03 stems from the fact that 1) the
mean luminosity of the LCS03 sample would lie in our high-redshift bin
($-20.5<M_B<-21.5$), and 2) LCS03 compare the oxygen abundance average
(weighted by the H$\beta$ luminosity) of relatively luminous
$M_B<-19.5$ CFRS galaxies with an average of NFGS galaxies spanning a
range of luminosities from $M_B=-16$ to $M_B=-22$.  The inclusion of
these low luminosity galaxies in the local mean reduces the
metallicity difference between the local and distant samples.

\subsection{The Effects of Sample Selection }

\begin{figure}
\plotone{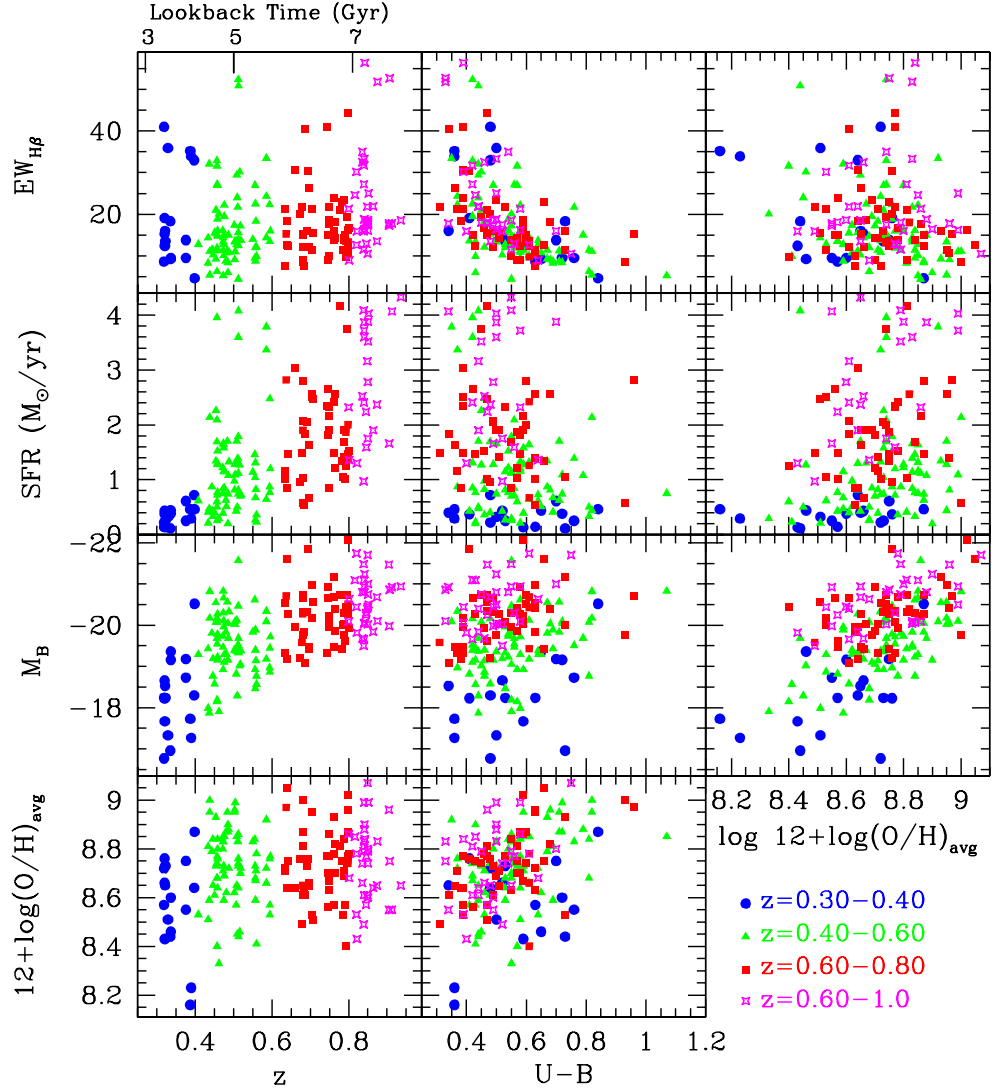}
\figcaption[multi] 
{Distribution of $H\beta$ equivalent width, star formation rate,
absolute B-band magnitude, oxygen abundance, U-B color, and redshift
for the \numselect\ selected TKRS galaxies.  
\label{multi} }
\end{figure}

\begin{figure}
\plotone{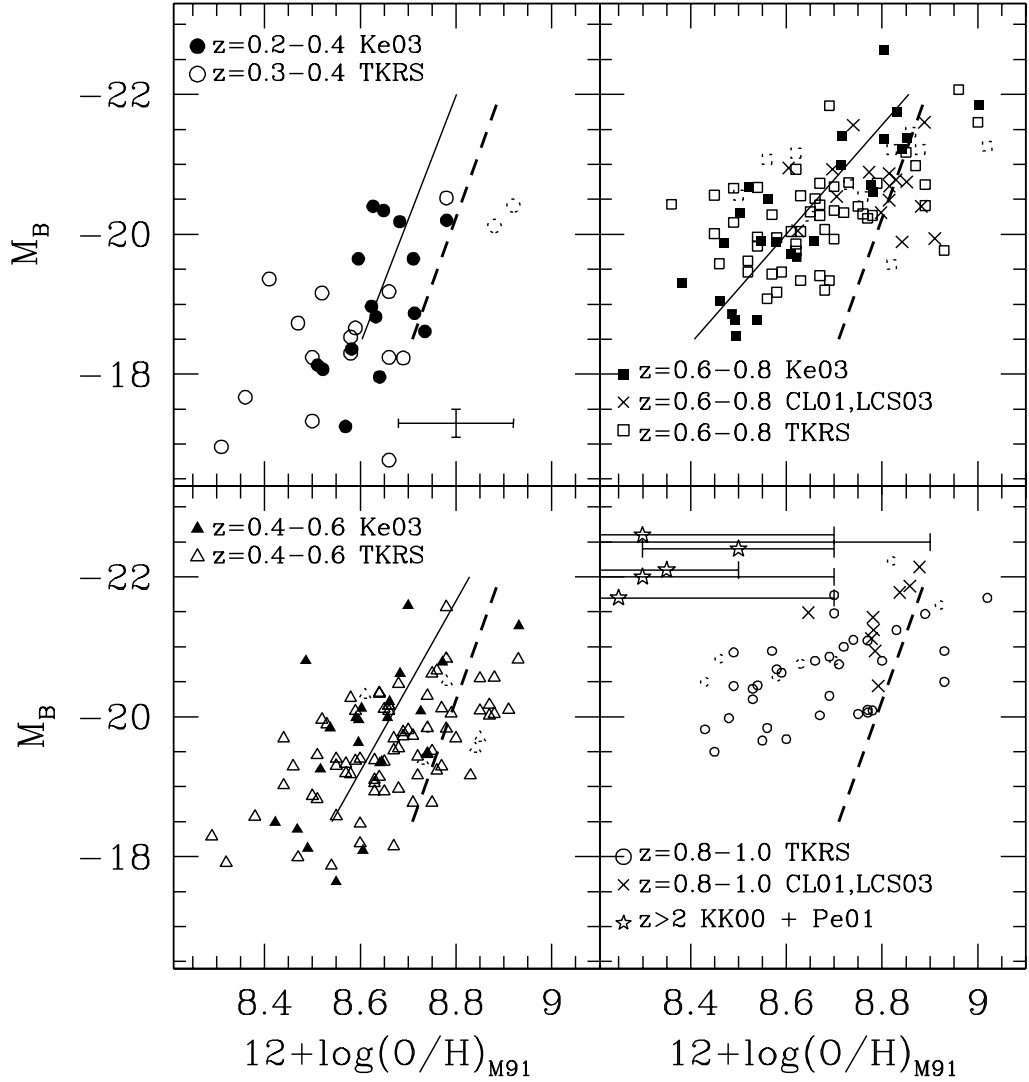}
\figcaption[LZ1] 
{Relation between blue luminosity and oxygen abundance in four
redshift bins for the TKRS galaxies in the GOODS-North field (open symbols),
the DEEP Groth Strip Survey from Kobulnicky
\etal\ (DGSS, 2003; filled symbols), and the Canada-France Redshift Survey from 
Lilly \etal\ (2003; crosses) and Corollo \& Lilly (2002; crosses).
Broken symbols denote 3$\sigma$ O/H lower limits for TKRS galaxies
with [O~III]$\lambda$5007 non-detections.  Solid lines are fits to the
DGSS data from Ke03 and dashed lines are fits to the local galaxies
samples defined in Ke03.  This figure shows oxygen abundances from
column~17 of Table~\ref{src.tab}.  Stars denote the high redshift
$z>2$ galaxies from Kobulnicky \& Koo (2000) and Pettini \etal\
(2001).  The TKRS data are consistent with the previous results from
smaller datasets.
\label{LZ1} }
\end{figure}

\begin{figure}
\plotone{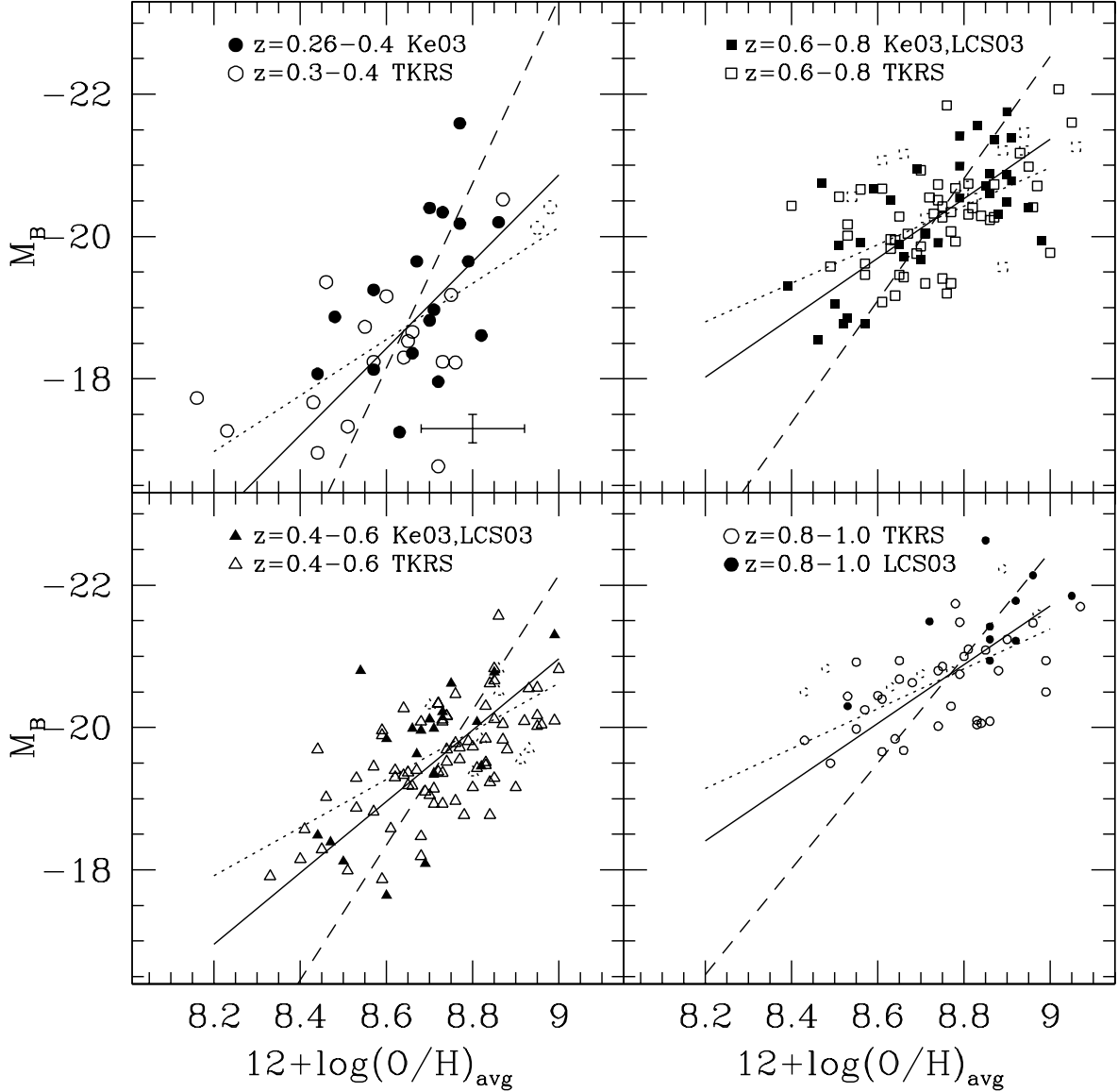}
\figcaption[LZ2] 
{Relation between blue luminosity and oxygen abundance as in
Figure~\ref{LZ1} except that oxygen abundances are $(O/H)_{avg}$ from
column 19 of Table~\ref{src.tab}.
Dotted and dashed lines are linear least squares fits for x-on-y and y-on-x
while the solid lines are linear bisectors of the two fits.
Parameters of the fits appear in Table~\ref{fit.tab}.  
\label{LZ2} }\end{figure}

\begin{figure}
\plotone{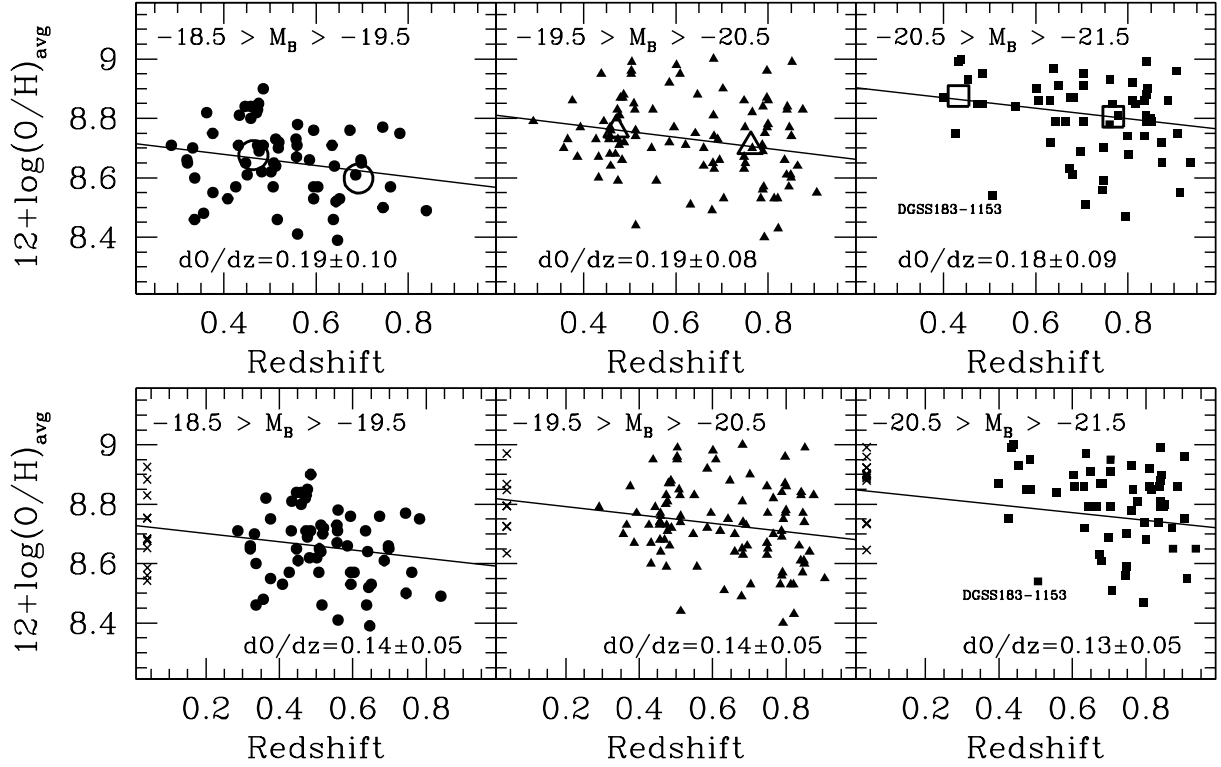}
\figcaption[zOH] 
{Relation between redshift and oxygen abundance for TKRS plus CFRS
(LCS03) and DGSS (Ke03) galaxies in
three different luminosity ranges.  The upper row
shows only distant galaxies.  Large open symbols indicate mean
values in two redshift bins: $z<0.6$ and $z>0.6$.  
The lower row shows the same data with
a sample of local $z=0$ galaxies from Jansen \etal\ (2001) and Kennicutt (1992) as 
defined by Ke03.  The slopes of least squares fits
are given in each panel.  The evolution of mean oxygen abundance
with redshift is $\sim0.19$ dex/z in the upper row and is consistent
across redshift bins.  When local galaxies are considered,
the slopes of the metallicity-redshift relations
in all bins drop to $\sim0.14\pm0.05$ dex/z.  
\label{zOH} }
\end{figure}

Could some selection effect in the chosen galaxy sample produce the
signature of evolution in the L-Z relation?  Kobulnicky \etal\ (2003)
discuss possible selection effects in the Deep Groth Strip Survey
sample and conclude that no identifiable selection effect can produce
the observed signature of chemo-luminous evolution with cosmic epoch.
Here we test the TKRS sample for selection effects which could mimic
the signature of genuine evolution in the L-Z relation.
Figure~\ref{LZtotal} shows the correlation between luminosity and
$12+\log(O/H)$ for 177 TKRS galaxies coded by redshift interval.  Dashed
lines are unweighted linear least squares fits of x-on-y and y-on-x,
while the solid line is the linear bisector of the two fits. This
figure shows that the highest redshift TKRS galaxies lie
systematically to the bright/metal-rich side of the overall L-Z
relation.  To examine whether some parameter other than redshift might
be responsible for this trend, we show in Figure~\ref{LZresid} the
magnitude residuals from the best fit L-Z relation (solid line in
Figure~\ref{LZtotal}) as a function of other fundamental galaxy
parameters: $M_B$, star formation rate, U-B color, redshift, and
ionization parameter indicator $O_{32}$.  There is no correlation
between magnitude residuals and $U-B$ or $O_{32}$.  There is a strong
correlation between magnitude residuals and $z$, SFR, and $M_B$.
These three parameters are not independent quantities.  As noted in
the discussion of Figure~\ref{multi}, $M_B$ is tightly coupled with
$SFR$ because it is used, along with $EW_{H\beta}$, to calculate the
star formation rate.  The $SFR$ scales with $M_B$.  Redshift and the
mean $M_B$ at a given redshift are also closely coupled by the
characteristics of a magnitude-limited survey.  Figures~\ref{scatter}
and \ref{multi} illustrate that galaxies as faint as $M_B=-17$
populate the lowest redshift bin while the highest redshift bin
contains no objects fainter than $M_B=-19.5$.  The lowest redshift bin
also lacks the population of luminous $M_B=-22$ galaxies found at
larger distances.  The correlations in Figure~\ref{LZresid}, then,
may all be understood as a consequence of a single underlying cause,
namely the inter-relation between $z$, $M_B$, and $SFR$ among sample
galaxies.

Which of the three parameters is the fundamental one driving the
observed change in the L-Z relation?  We argue that redshift is
fundamental.  Because the least squares fit in Figure~\ref{LZtotal}
uses $M_B$ as one of the correlative variables, residuals
should not depend on $M_B$.  If the star formation rate were the
fundamental parameter driving the evolution of the L-Z relation (i.e.,
galaxies with the highest star formation rates preferentially lie on
the bright/metal-poor side of the L-Z relation), then we would also
expect related parameters like the U-B color, which is sensitive to
the star formation rate, to show a correlation with $\Delta M_B$ as
well.  Figure~\ref{LZresid} shows that $U-B$ is not correlated with
  $\Delta M_B$.  We are left with the conclusion that
the evolution of the L-Z relation is driven primarily by the
redshift of the galaxies under consideration.  Galaxies of
a given luminosity are, on average, increasingly metal poor
at higher redshifts.  Said another way, galaxies of a given
metallicity are, on average, more luminous at higher redshifts.  
This effect is most pronounced among the least luminous galaxies,
those with $M_B$ fainter than $\sim-20$.  None of the identified
sample selection effects would produce the changes in the nature
of the L-Z relation seen in Figures~\ref{LZ2} and \ref{zOH}.

\subsection{Comparison to Metallicity Evolution in the Milky Way }

One goal of studying galaxy populations at cosmological distances is
to understand the evolutionary paths of individual galaxies.  However,
observing the evolution of the mean chemo-luminous properties of the
population of star-forming galaxies over the last 8 Gyr
(Figure~\ref{LZ2}, \ref{zOH}) is not the same thing as observing the
evolution of any particular galaxy.  Events shaping the evolution of
any given galaxy at any particular point in its history, could, in
principle, be unique to that galaxy and not be reflected in the mean
properties of galaxies at the equivalent lookback time in the distant
universe.  A galaxy, may, for example, spend most of its existence in
a quiescent passively evolving phase after a brief initial period of
star formation.  Other galaxies might form stars continuously
throughout their existence, while still others may not begin the star
formation process until comparatively recent times.  However, if the
boundary conditions for galaxy evolution are determined primarily by
global cosmological parameters, such as the age of the universe, the
expansion rate, and the density of dark matter and gas available to
form stars, then evolutionary paths of most galaxies ought to 
be observable in the mean evolution of galaxy properties with redshift.

\begin{figure}
\plotone{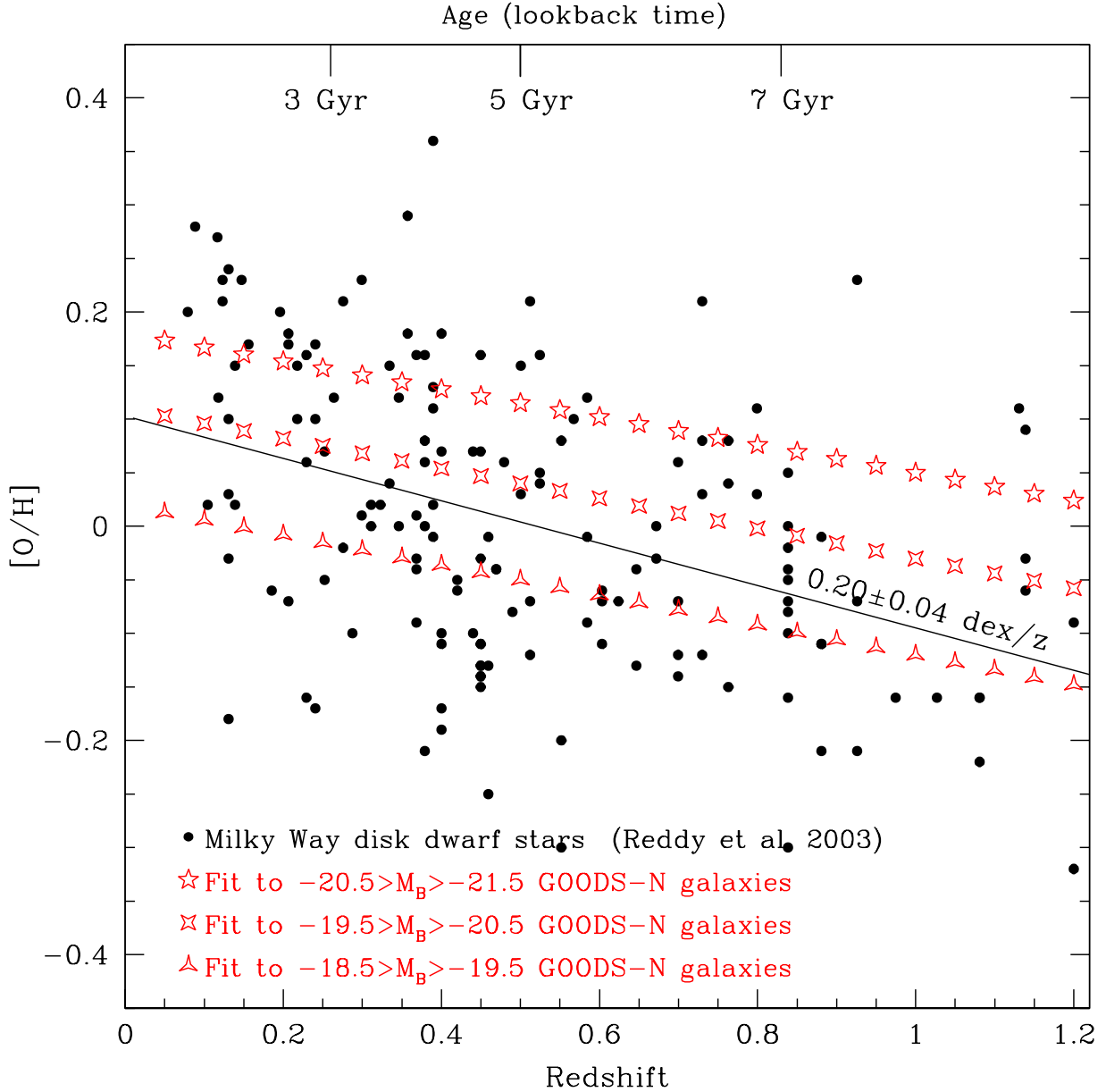}
\figcaption[zOHMW] 
{Relation between age (a.k.a lookback time) and oxygen abundance,
relative to solar, for Milky Way disk stars from Reddy \etal\ (2003).
The ages of the stars are plotted in terms of their equivalent
redshifts (using the adopted cosmology) for direct comparison to
Figure~\ref{zOH}.  The line shows a least squares linear
fit to the Galactic data.  
The tracks of symbols show the fits from Figure~\ref{zOH} (lower row)
for galaxies in three different luminosity bins. 
The mean slope and zero point for the evolution of Milky Way stars
are in good agreement with the overall metallicity and
rate of enrichment for distant galaxies.  
\label{zOHMW} }
\end{figure}

\begin{figure}
\plotone{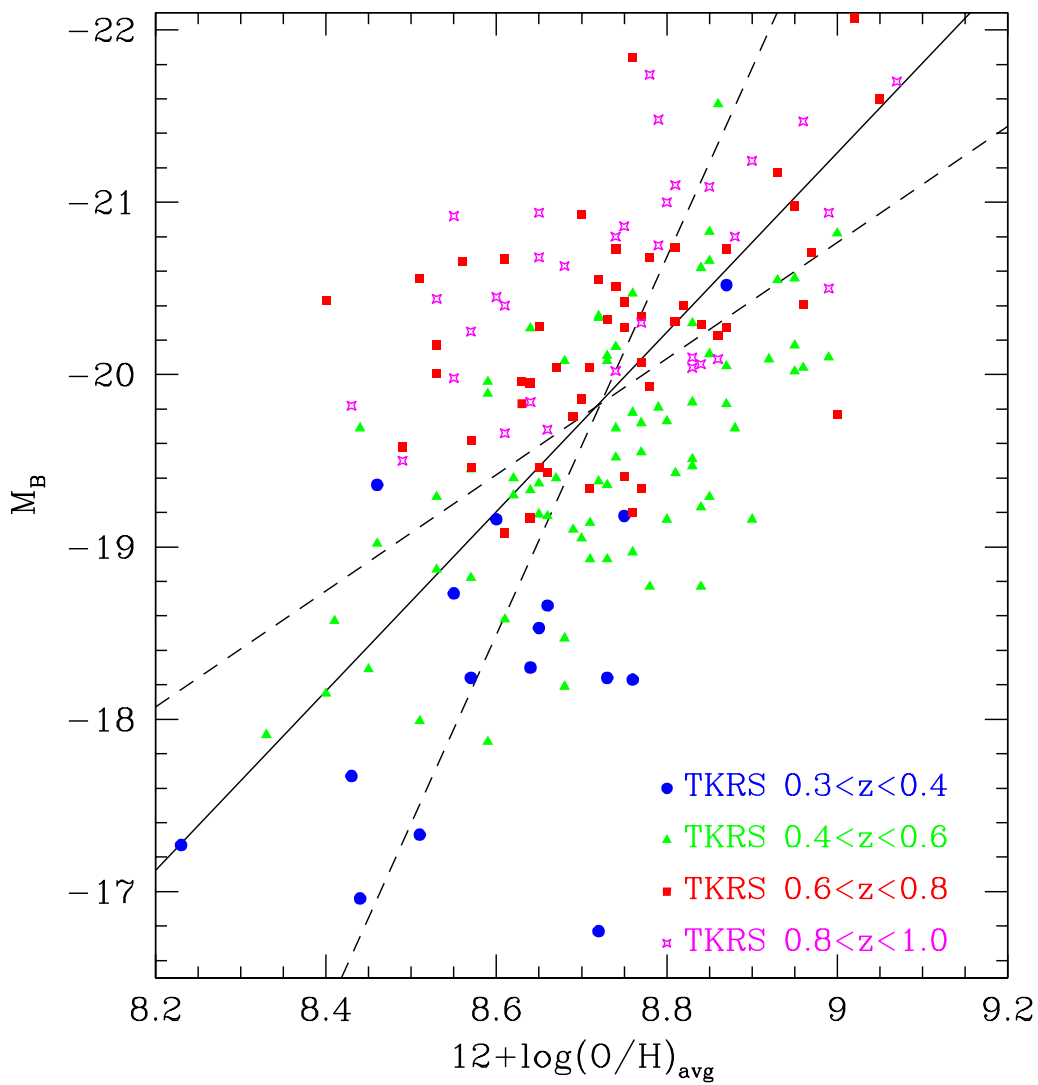}
\figcaption[LZtotal] 
{L-Z relation for 177 TKRS galaxies with symbols denoting the four 
redshift bins.  Dashed lines are unweighted linear least squares fits
of x-on-y and y-on-x, while the solid line is the linear bisector
of the two fits.  This figure shows that the highest redshift TKRS galaxies
lie systematically to the bright/metal-rich side of the overall L-Z relation.
\label{LZtotal} }
\end{figure}

\begin{figure}
\plotone{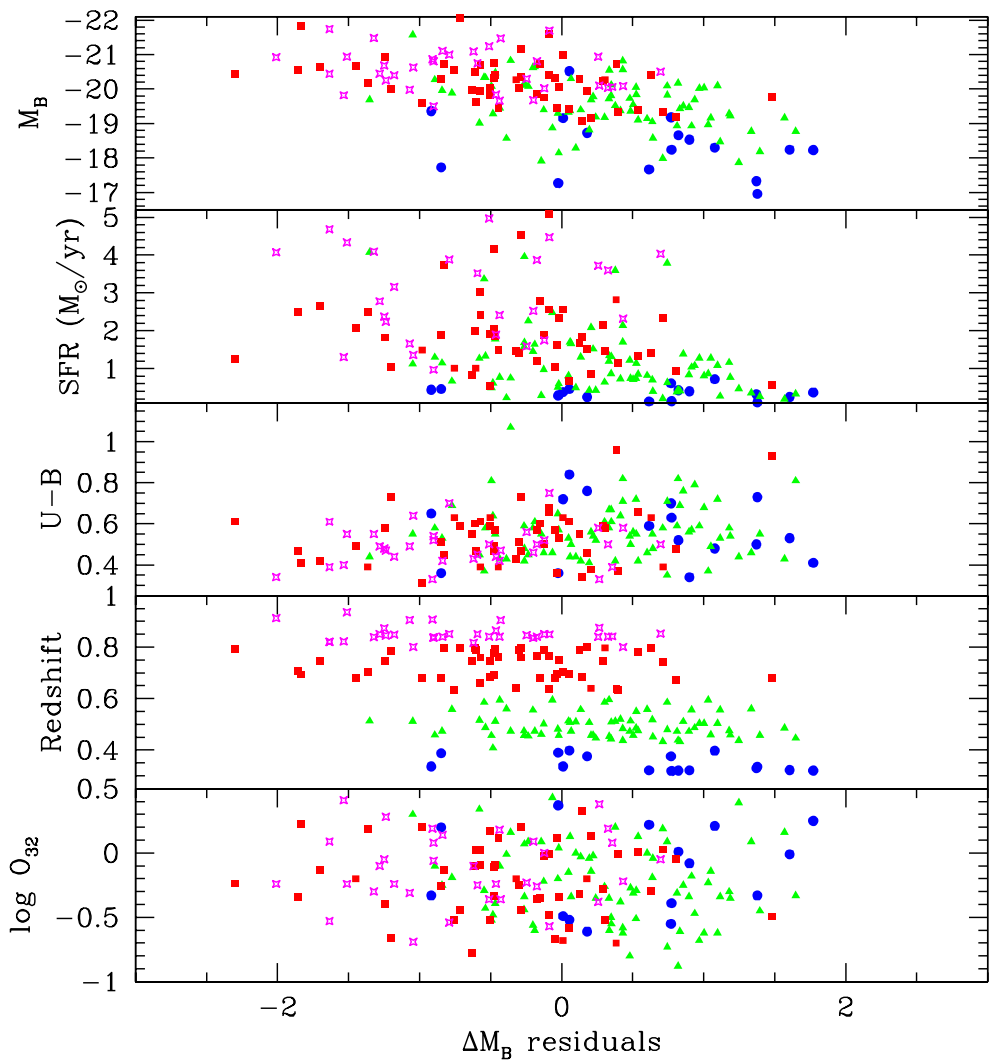}
\figcaption[LZresid] 
{Magnitude residuals, $\Delta M_B$, from the best-fit luminosity-metallicity 
relation in Figure~\ref{LZtotal} for 177 TKRS galaxies as a function of
luminosity, star formation rate, color, redshift, and ionization parameter
indicator, $O_{32}$.  Symbols denote redshift bins.   
There is a strong correlation between
magnitude residuals and the three closely related parameters 
$z$, $SFR$, and $M_B$.
There is no correlation between residuals and U-B or $O_{32}$. 
\label{LZresid} }
\end{figure}

Measuring ages and chemical abundances of stars in the Milky Way
provides a glimpse into the history of one, presumably typical, disk
galaxy.  Figure~\ref{zOHMW} shows [O/H], the logarithmic oxygen
abundances relative to solar, of F and G stars in the Galactic disk as
a function of age (Reddy \etal\ 2003).  We have plotted on the
ordinate the redshift corresponding to the lookback time of the
measured age of the star.  For our adopted cosmology of $H_0=70$ km
s$^{-1}$ Mpc$^{-1}$, $\Omega_M=0.3$, $\Omega_\Lambda=0.7$, the
relation between redshift, $z$, and the lookback time or age, $a$, in
Gyr is closely approximated by a polynomial,

\begin{equation}
z=1.3356-2.0847a + 1.2431a^2 -0.33620a^3+0.046737a^4-
0.0032106a^5+0.000086938a^6.
\end{equation}

\noindent The solid line shows the least squares O/H on $z$ fit
to the stellar measurements.  The tracks of starred symbols show the
least-squares linear fits to distant GOODS-N+CFRS+TKRS  
galaxies in the three luminosity
bins from the lower row of Figure~\ref{zOH}.  
We use [O/H] = 12+log(O/H) -8.7 in
keeping with the most recent solar oxygen abundance measurements
(Allende-Prieto \etal\ 2001).  Both the overall level of metal
enrichment (zero point) for $M_B\sim20$ galaxies and the rate at which
oxygen abundance increases with time (slope) among all GOODS-N
galaxies over the redshift range $z$=1--0 agree well with the trend
observed in Galactic stars.  The slopes of the best fit relations for
galaxies from the lower panel of Figure~\ref{zOH}, $0.14\pm0.05$ dex, are
within the uncertainties of the slope for Milky Way stars,
$0.20\pm0.04$ dex over the range $0.0<z<1.0$.

LCS03 presented a diagram similar to Figure~\ref{zOHMW} where they
compare CFRS and NFGS oxygen abundances to [Fe/H] measurements of
Galactic disk stars and noted that the stars were consistently 0.2-0.3
dex more metal poor than the galaxies.  This offset may be understood
as the signature of super-solar O/Fe ratios found in Galactic stars
over most of the history of the Milky Way reflecting varying
nucleosynthetic sources (reviewed by Wheeler, Sneden, \& Truran 1989).
Using oxygen measurements for Milky Way stars in Figure~\ref{zOHMW}
shows good agreement with the oxygen abundance measurements in distant
galaxies.  While the luminosity history of the Milky Way is not known,
and while the dispersion in oxygen abundances for Galactic stars is
large, the chemical enrichment process that occurred in the disk of
the Milky Way appears to be a good representation of the chemical
enrichment process in the bulk of the star-forming galaxies over the
last 8 Gyr. Ke03 observed that the majority of the star--forming
galaxies in the DEEP Groth Strip Survey over a similar redshift range
appeared to have substantial disk components based on Hubble Space
Telescope imaging.  Thus, it appears we are able to observe directly,
in ensembles of disk-like galaxies at cosmological distances, the same
chemical histories encoded in Galactic stellar populations.  In
general, we expect that the luminosity-weighted nebular oxygen
abundance of the entire Milky Way would be 0.1--0.3 dex higher than
the metallicity of the solar neighborhood, given that the bulk of star
formation occurs in the molecular ring at smaller radii where the
average composition is more metal rich (e.g., Shaver \etal\ 1983;
Maciel, DaCosta, \& Uchida 2003).

For a galaxy that evolves as a ``closed box'' (i.e., no
gas inflow or outflow),
converting gas to stars with a fixed initial mass function and
chemical yield, the metallicity is determined by a single parameter:
the gas mass fraction, $\mu=M_{gas}/(M_{gas}+M_{stars})$.  The
metallicity, $Z$, is the ratio of mass in elements heavier than He to
the total mass and is given by

\begin{equation}
Z= Y~ln(1/\mu) \label{mu},
\end{equation}

\noindent where $Y$ is the ``yield'' as a mass fraction.  A typical
total metal yield for a Salpeter IMF integrated over 0.2--100
$M_\odot$ is $Y=0.012$ by mass (i.e., 2/3 the solar metallicity of
0.018; see Pagel 1997, Chapter 8). A total oxygen yield for the same
IMF would be $Y_{O}=0.006$.  Effective yields in many local galaxies
range from solar to factors of several lower (Kennicutt \& Skillman
2001; Garnett 2002).  The change in metallicity with gas mass fraction is 
independent of yield and is given by

\begin{equation}
{{d\log Z}\over{d\mu}} = {{d}\over{d\mu}}~\big[Y~ln({{1}\over{\mu}})\big]
  = {{0.434}\over{\mu~ln(\mu)}}.    \label{delta}
\end{equation}

\noindent For reasonable values of the gas mass fraction (0.1-0.3), 
the change in gas mass fraction, $\Delta\mu$, corresponding to $\Delta
\log(Z)=0.14$ is 0.07 to 0.12.  Such a change in average gas content
should be observable with future radio wave interferometers.

\subsection{Comparison to Expectations of Cosmic Star Formation Models }

The observed 0.14 dex increase in oxygen abundance of galaxies from
$z=1$ to $z=0$ is equivalent to a 38\% increase in metallicity since
$z=1$. In other words, 28\% of the metals in $z=0$ star-forming
galaxies $-18.5<M_B<20.5$ have been produced in the last 7.7 Gyr since
$z=1$.  This conclusion can be tested for consistency against
expectations from the global rate of cosmic star formation over the
same period.  Given that the magnitude range of our sample encompasses
the bulk of luminosity produced by all galaxies (given a typical local
galaxy luminosity function), the observed 28\% increase in metal
content should reflect the overall level of chemical enrichment in the
universe.  The rate of metal enrichment should correlate directly with
the rate of star formation,
subject to the condition that the stellar initial mass function
and the chemical yield per mass of stars formed is constant.

\begin{figure}
\plotone{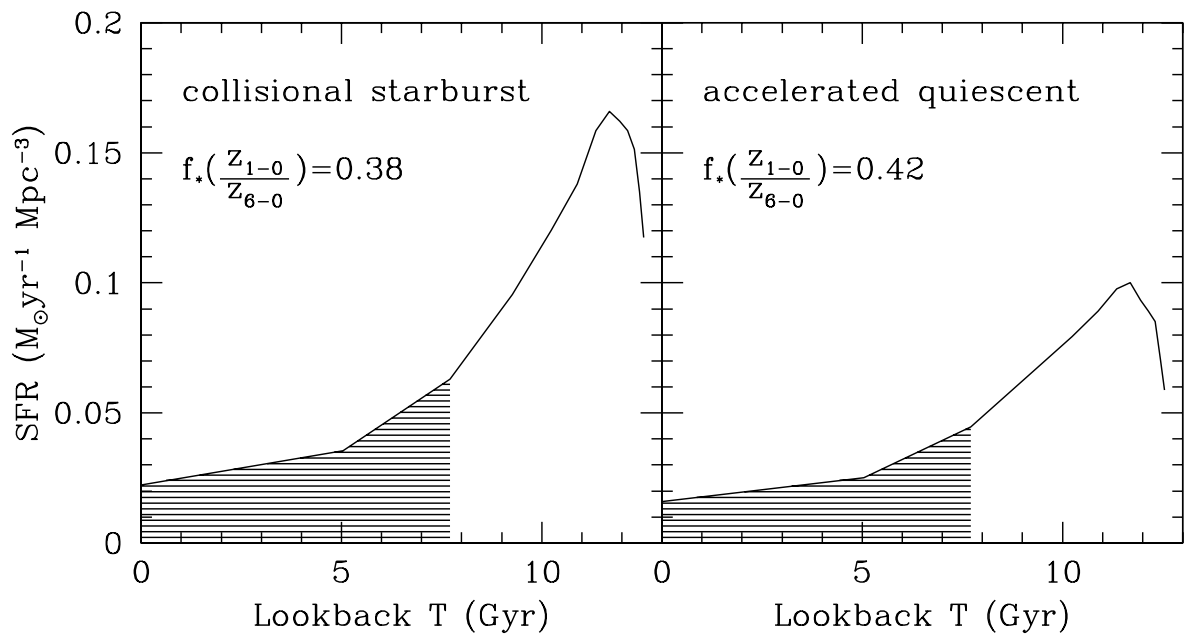}
\figcaption[model] 
{Star formation rate density as a function of time, for our adopted cosmolgy,
based on two models from Figure~9 of Somerville \etal\ (2001).  The shaded
regions in each figure show the integral of the star formation rate
over the period from $z=0$ to $z=1$ (0 Gyr to 7.7 Gyr lookback time).
The fraction of all stars formed in the last 7.7 Gyr, $f_*({{z1-0}\over{z6-0}})$, 
is 0.38 and 0.42 respectively, for the
``collisional starburst'' and ``accelerated quiescent'' models. 
\label{model} }
\end{figure}

The cosmic rate of star formation as a function of redshift has been
extensively studied, and we adopt, for illustrative purposes, the
models from Figure~9 of Somerville \etal\ (2001).  Figure~\ref{model}
shows the star formation rate versus time for their ``collisional
starburst'' and ``accelerated quiescent'' models which
give reasonable agreement with the observations.  We have transformed
redshift into linear time on the ordinate, using our adopted
cosmology, and plotted linear star formation rate on the abscissa.
The shaded regions in each figure show the integral of the star
formation rate over time from $z=0$ to $z=1$ (0 Gyr to 7.7 Gyr
lookback time).  The fraction of all stars formed in the last 7.7 Gyr,
$f_*({{z1-0}\over{z6-0}})$, is 0.38 and 0.42 in the two models.
This fraction compares favorably with the fraction of
metals formed during the same time period, $0.28\pm0.07$, from 
the previous section.
Any difference, if real,  between the predicted fraction based on
star formation rate indicators and the observed
fraction may be explained in several ways.

\begin{itemize}
\item{The present data/models 
may be underestimating the fraction of star formation which occurred
before $z=1$.  A star formation rate which is, on average, not more
than 10\% higher at $z>1$ would be required to produce agreement.}
\item{The stellar initial mass function or metal yield may vary with
redshift so that the assumption of a linear relationship between star
and metal production is violated.  A stellar initial mass function
which is more shallow or more top heavy, or a higher effective
nucleosynthetic yield at $z>1$ would be required.}
\end{itemize}
  
Given the significant uncertainties on the star formation rate and the
metal enrichment rate in galaxies as a function of redshift, no firm
conclusion can be drawn except to say that the two rates are
generally consistent with one another.
Given that all models and data on the star
formation rate (e.g., see Figure~\ref{model} and the summary in
Somerville \etal\ 2001) show a decline in the star formation rate with
time after $z=1$, the rate of metal enrichment in galaxies should also
drop.  In principle, this drop in the rate of metal enrichment should
be observable in chemical studies of galaxies at cosmological
distances, but will require a level of measurement precision and/or a
sample size beyond current capabilities.

\section{Discussion}

Our findings with the GOODS-N data here are in agreement with the
conclusions of Ke03, namely that chemo-luminous evolution is most
pronounced among the least luminous (and possibly the least massive)
galaxies during the 8 Gyr since $z\sim1$.  The change in slope of
the L-Z relation with redshift in Figure~\ref{LZ2} is due to the
emergence of a population of moderately luminous ($M_B\sim-20$) galaxies with
intermediate metallicities (12+log(O/H)$\sim8.5-8.6$) at redshifts
beyond $z=0.6$ which are not seen in local samples.  This observation
is consistent with the conclusions of LCS03 who advocate a progression
of star formation activity from massive galaxies to less massive
galaxies with decreasing redshift, a process generically termed
as ``downsizing'' (Cowie \etal\ 1996). In the context of single-zone
\textsc{P\'egase2} galaxy evolution models, Kobulnicky \etal\ (2003),
concluded that the change in slope of the L-Z relation could be
explained by at least two of the following three phenomena: 1)
low-mass galaxies have lower effective chemical yields than massive
galaxies, 2) low-mass galaxies assemble on longer timescales than
massive galaxies, 3) low-mass galaxies began the assembly process at a
later epoch than massive galaxies, i.e., ``downsizing''. 
The possibility that low mass
galaxies begin their assembly at a later cosmic epoch has received
several independent sources of support, both theoretical and
observational.  Babul \& Rees (1992) proposed a theoretical model
whereby photoionization from the first generations of cosmic star
formation keeps gas in galaxies with small potential wells ionized
(and thus unable to form stars) until some relatively late epoch,
approximately $z=1$.  Skillman \etal\ (2003) concluded from an Hubble
Space Telescope study of the star formation history in the Local Group
galaxy IC 1613 that its star formation may have been inhibited until
$z\sim1$.  Kodama \etal\ (2004) concluded from color-magnitude relations
that galaxies in the
Subaru/XMM Deep Survey were consistent with ``downsizing'' scenario as
well.
 
Can the evidence for the ``downsizing'' scenario be predicted or
modeled theoretically?  Theoretical hydrodynamic simulations have been
used recently to predict the variation of metallicity with redshift
for Damped Lyman $\alpha$ (DLA) absorbers (e.g., Nagamine, Springel,
\& Hernquist 2004) or for the stellar metallicity of galaxies 
(e.g., Nagamine, Fukugita, \& Ostriker 2001).  Unfortunately, the
current models suffer from a lack of resolution in the $0<z<1$ range.
The direct simulations by Nagamine \etal\ predict that the mean
metallicity at $z=1$ is $\sim 0.4$ \Zsun, corresponding to
\OH$\sim8.3$, significantly lower than the metallicities of our
sample, even at $z$ close to 1.  However, both absorption line
measurements and simulations of chemical abundances in DLA systems are
not directly comparable to emission line measurements because the
former probe the more extended gaseous halos surrounding galaxies
while the latter probe the \HII\ regions and sites of active star
formation within the inner disks of galaxies.  These chemical
measurements of galaxies in the GOODS-N provide a significant dataset
for comparison with future cosmological simulations which will have
the temporal and spatial resolution to track the composition of the
galaxies on 10 kpc scales with redshift.

\section{Conclusions}

We have parameterized an updated analytic formulation of the
$R_{23}$--O/H relations for estimating nebular oxygen abundances based
on the photionization models of Kewley \& Dopita (2003). After
reviewing existing calibrations, we also provide a parameterization
for the average of this calibration and that of McGaugh (1991) for the
upper branch only.  An additional parameterization may be used to
estimate, albeit very crudely, the metallicities of galaxies based on
the nebular [N~II]/H$\alpha$ ratios.

Analyzing spectra of \numselect\ galaxies at $0.3<z<1.0$ 
from the GOODS-N TKRS, we measure
galaxy-averaged nebular oxygen abundances of $8.2<12+\log(O/H)<9.1$,
corresponding to metallicities between 0.3 and 2.5 times the solar
value.  The overall oxygen abundance of galaxies in the luminosity
range $-18.5<M_B<-21.5$ increases by 0.14$\pm$0.05 dex from $z=1$ to
$z=0$.  Said another way, galaxies in this intermediate-redshift
sample are 1-3 magnitudes
more luminous at a given metallicity than are local counterparts.  For
closed box chemical evolution models, the implied change in gas mass
fraction, $\mu$, over the $z=$1--0 interval as gas is cycled through
stars to produce heavy elements is $\Delta\mu\sim0.10$.  This sample
of galaxies exhibits a luminosity-metallicity correlation, but with
different zero points, and possibly different slopes at each redshift
interval.  The change in slope is driven mostly by the appearance of a
population of moderate luminosity ($M_B\sim-20$) galaxies at $z>0.6$
with intermediate metallicities (12+log(O/H)=8.5-8.6).  This population
is likely to evolve into the comparatively luminous, metal rich disk
galaxy population of today.  This change in
galaxy populations is consistent with a later formation epoch for
lower mass galaxies.   The increase in the
mean oxygen abundance of $-18.5<M_B<-21.5$ galaxies is broadly
consistent with the global picture of cosmic star formation activity
which suggests that $\sim38$\% of the stars and $\sim28\pm0.07$\% of
the metals in the universe have formed in the 7.7 Gyr since $z=1$.

\acknowledgments

We thank David Koo for scientific inspiration and Christopher Willmer
for use of his K-correction code.  We are grateful to the TKRS Team,
the W.~M. Keck Observatory Director, and the DEEP2 Redshift Survey
Team for making these data possible and for making them available in a
timely fashion to the larger astronomical community.  We thank Danny
Dale, Christy Tremonti, Alice Shapley, and Henry Lee 
for helpful conversations.  Simon Lilly, the
referee, provided key insights which improved the clarity and
completeness of this paper.  H.~A.~K was supported by NASA through
NRA-00-01-LTSA-052.  L.~J.~K. was supported by a Harvard-Smithsonian
CfA Fellowship.

\clearpage

\clearpage

\clearpage

\epsscale{0.7}

\clearpage

\epsscale{1.0}

\clearpage

\clearpage

\clearpage

\clearpage

\clearpage

\clearpage

\clearpage

\clearpage

\clearpage

\clearpage

\clearpage

\clearpage

\begin{deluxetable}{rrrrccccccccccccccccc}

\rotate
\tabletypesize{\tiny} 
\setlength{\tabcolsep}{0.01in} 
\tablewidth{8in}
\tablecaption{GOOD-North TKRS Selected Galaxies \label{src.tab}}
\tablehead{
\colhead{\ } & 
\colhead{ID} & 
\colhead{GOODS ID} & 
\colhead{z}   &
\colhead{${R} $} & 
\colhead{$i_{775W}$} & 
\colhead{$M_B$} & 
\colhead{$U-B$} &
\colhead{$EW$}   &
\colhead{$EW$}   &
\colhead{$EW$}  &
\colhead{$EW$}  &
\colhead{$SFR$} &
\colhead{code} &
\colhead{$log(O_{32})$} &
\colhead{$R_{23}$} &
\colhead{$O/H$}  &
\colhead{$O/H$}  &
\colhead{$O/H$}  &
\colhead{$O/H$}  &
\colhead{$O/H$}    \\ [.2ex]
\colhead{\ } & 
\colhead{} & 
\colhead{} & 
\colhead{}   &
\colhead{mag} & 
\colhead{mag} & 
\colhead{mag} & 
\colhead{mag} &
\colhead{$[O~II]$}   &
\colhead{$H\beta$}   &
\colhead{$[O~III]$}  &
\colhead{$ _{H\alpha/N\sc{II}}$}  &
\colhead{$M_\odot/yr$} &
\colhead{} &
\colhead{} &
\colhead{} &
\colhead{$M91$}  &
\colhead{$KD$}  &
\colhead{$avg$}  &
\colhead{$N\sc{II}$}  &
\colhead{$final$}    \\ [.2ex]
\colhead{(1)} &  
\colhead{(2)} &  
\colhead{(3)} &  
\colhead{(4)} &  
\colhead{(5)} &
\colhead{(6)} &
\colhead{(7)} &
\colhead{(8)} &
\colhead{(9)} &
\colhead{(10)} &
\colhead{(11)} &
\colhead{(12)} &
\colhead{(13)} &
\colhead{(14)} &
\colhead{(15)} &
\colhead{(16)}  &
\colhead{(17)}  &
\colhead{(18)}  &
\colhead{(19)}  &
\colhead{(20)}  & 
\colhead{(21)}    } 
\startdata
  1 &   3551 &  J123651.06+621732.0 &  0.31870 &  22.57 &  22.12 & -18.24 & 0.630 &    40.9$\pm$2.1 &    6.6$\pm$ 0.5 &   16.5$\pm$ 0.7 &          7.1 & 0.14 &  0 &  -0.39 &   6.67  &  8.5$\pm$0.04 &    8.65 &    8.57  &    8.63  &   8.57$\pm$0.15 \\
  2 &   3792 &  J123656.21+621743.2 &  0.31911 &  24.21 &  24.00 & -16.77 & 0.480 &    79.4$\pm$6.5 &     39$\pm$ 1.9 &  157.8$\pm$ 1.8 &      \nodata & 0.22 &  0 &   0.29 &   5.78  & 8.66$\pm$0.02 &    8.79 &    8.72  & \nodata  &   8.72$\pm$0.15 \\
  3 &   9483 &  J123652.47+621036.0 &  0.32036 &  22.18 &  21.89 & -18.66 & 0.520 &    46.7$\pm$4.9 &   13.3$\pm$ 1.3 &   47.9$\pm$ 1.5 &        10.43 & 0.43 &  0 &   0.01 &   6.18  & 8.59$\pm$0.04 &    8.73 &    8.66  &    8.62  &   8.66$\pm$0.15 \\
  4 &   7972 &  J123658.06+621300.8 &  0.32040 &  22.73 &  22.45 & -18.23 & 0.410 &    36.7$\pm$3.4 &   17.1$\pm$ 1.3 &   65.8$\pm$ 1.7 &        17.35 & 0.37 &  0 &   0.25 &   5.36  & 8.69$\pm$0.02 &    8.83 &    8.76  &    8.46  &   8.76$\pm$0.15 \\
  5 &  11168 &  J123704.28+621000.0 &  0.32129 &  23.35 &  23.07 & -17.67 & 0.590 &    45.2$\pm$7.4 &   10.4$\pm$ 1.2 &   75.3$\pm$ 1.5 &      $>$7.16 & 0.13 &  0 &   0.22 &   9.71  & 8.36$\pm$0.07 &    8.51 &    8.43  & $<$8.96  &   8.43$\pm$0.16 \\
  6 &  11004 &  J123703.91+621009.7 &  0.32144 &  22.45 &  22.21 & -18.53 & 0.340 &    54.4$\pm$3.1 &     14$\pm$ 1.1 &   45.1$\pm$ 1.4 &         $>$9 &  0.4 &  0 &  -0.08 &   6.21  & 8.58$\pm$0.03 &    8.72 &    8.65  & $<$8.65  &   8.65$\pm$0.15 \\
  7 &  12136 &  J123744.51+621411.1 &  0.32238 &  22.62 &  22.28 & -18.24 & 0.530 &    37.7$\pm$3.8 &   11.5$\pm$ 1.0 &   36.3$\pm$ 1.3 &        10.46 & 0.25 &  0 &  -0.01 &   5.48  & 8.66$\pm$0.04 &     8.8 &    8.73  &    8.61  &   8.73$\pm$0.15 \\
\enddata
\end{deluxetable}

\newpage
\begin{deluxetable}{clrrrrrrr}
\tablewidth{4.3in}
\tabletypesize{\scriptsize}
\tablecaption{$O_{32}$ and $R_{32}$ values for curves shown in Figure~\ref{R23_vs_O32} \label{R23_O32.tab}}
\tablehead{ 
$q$     & line & \multicolumn{7}{c}{Z} \\
(cm/s)  & ratio & 0.05 & 0.1 & 0.2 & 0.5 & 1.0 & 1.5 & 2.0 \\}
\startdata
$5\times10^{6}$ & $\log(O_{32})$ & -1.03 & -1.21 & -1.27 & -1.43 & -1.59 & -1.72 & -1.87 \\
                & $\log(R_{23})$ &  0.04 & 6.67 & 8.76 & 2.60 & 2.26 & 5.54 & 5.57 \\
$1\times10^7$ & $\log(O_{32})$ &  -0.51 & -0.61 & -0.76 & -0.87 & -1.09 & -1.10 & -1.33 \\
 &              $\log(R_{23})$ &   0.16 & 5.79 & 8.99 & 6.62 & 2.76 & 5.47 & 4.0 \\
$2\times10^7$ & $\log(O_{32})$ &  -0.00 & -0.11 & -0.26 & -0.41 & -0.63 & -0.68 & -0.90 \\
              & $\log(R_{23})$ &  0.27 & 6.18 & 7.91 & 3.72 & 4.50 & 5.63 & 6.28 \\
$4\times10^7$ & $\log(O_{32})$ &  0.44 &  0.33 &  0.17 & -0.02 & -0.25 & -0.34 & -0.57 \\
 &              $\log(R_{23})$ &  0.38 & 5.37 & 6.36 &  7.01 & 4.10 & 5.11 & 5.10 \\
$8\times10^7$  & $\log(O_{32})$ &  0.85 & 0.72 & 0.55 & 0.33 & 0.07 & -0.09 & -0.37 \\
               & $\log(R_{23})$ &  0.48 & 9.72 & 4.80 & 4.25 & 6.01 & 5.19 & 4.59 \\
$1.5\times10^8$ & $\log(O_{32})$ &  1.19 & 1.06 & 0.87 & 0.62 & 0.32 & 0.06 & -0.31 \\
                 & $\log(R_{23})$ &  0.56 & 6.22 & 6.73 & 8.04 & 6.68 & 5.30 & 9.91 \\
\enddata
\end{deluxetable}

\begin{deluxetable}{lcccc}
\tabletypesize{\scriptsize} 
\setlength{\tabcolsep}{0.02in} 
\tablewidth{4.3in}
\tablecaption{L-Z Relation Fits: $12+log(O/H)_{avg}=X\times M_B + Y $ \label{fit.tab}}
\tablehead{
\colhead{Sample} & 
\colhead{z} & 
\colhead{\#Galaxies} & 
\colhead{$X_{avg}$}   & 
\colhead{$Y_{avg}$}    
  \\ [.2ex]
\colhead{(1)} & 
\colhead{(2)} & 
\colhead{(3)} & 
\colhead{(4)} & 
\colhead{(5)}  }
\startdata
O/H on $M_B$ fits \\
DGSS+CFRS+TKRS & 0.2-0.4   & 32  & -0.077$\pm$0.032 &  7.20   \\
DGSS+CFRS+TKRS & 0.4-0.6   & 97  & -0.105$\pm$0.017 &  6.66   \\
DGSS+CFRS+TKRS & 0.6-0.8   & 84  & -0.117$\pm$0.017 &  6.37   \\
DGSS+CFRS+TKRS & 0.8-1.0   & 44  & -0.134$\pm$0.032 &  5.97   \\
Linear bisector fits \\
DGSS+CFRS+TKRS & 0.2-0.4   & 32  & -0.164$\pm$0.026 &  5.57   \\
DGSS+CFRS+TKRS & 0.4-0.6   & 97  & -0.199$\pm$0.016 &  4.82   \\
DGSS+CFRS+TKRS & 0.6-0.8   & 84  & -0.239$\pm$0.029 &  3.88   \\
DGSS+CFRS+TKRS & 0.8-1.0   & 44  & -0.241$\pm$0.027 &  3.73   \\\enddata
\tablerefs{
(1) galaxy sample, by redshift;
(2) redshift range;
(3) number of galaxies in sample;
(4) slope of metallicity-luminosity relation and uncertainty;
(5) intercept of the least squares fit.}
\end{deluxetable}


\begin{references}

\reference{} Anders, E., \& Grevesse, N. 1989, GeCoA, 53, 197 
\reference{} Babul, A., \& Rees, M.~J. 1992, MNRAS 255, 346
\reference{} Bruzual, G. \& Charlot, S. 2003, \mnras, 344, 1000 
\reference{} Carollo, C., M. \& Lilly, S.~J. 2001, ApJ, 548, L153 (CL01)
\reference{} Cowie, L.~L., Songalia, A.~A., Hu, E.~M., Cohen, J.~G., 1996, AJ, 112, 839 
\reference{} Davis, M. \etal\ (The DEEP2 Team) 2003, SPIE 4834, 161
\reference{} Dopita, M.~A., \& Evans, I.~N. 1986, ApJ, 307, 431
\reference{}  Dopita, M. A., Kewley, L. J.,
Heisler, C. A., \& Sutherland, R. S. 2000, \apj, 542,224
\reference{} van Dokkum, P. \etal\ 2004, \apj, {\it in press}, astro-ph/0404471
\reference{} Edmunds, M.~G. \& Pagel, B.~E.~J. 1984, MNRAS, 211, 507
\reference{} Faber, S.~M. 1973, ApJ, 179, 423
\reference{} Faber, S.~M. \etal\ 2003, SPIE, 4841, 1657
\reference{} Ferland, G. J. \& Truran, J. W. 1981, \apj, 244, 1022
\reference{} Fioc, M. \& Rocca-$\!$Volmerange, B. 1999, astro-ph/9912179 
\reference{} Garnett, D.~R., Kennicutt, R.~C. \& Bresolin, F. 2004, ApJ, 607, 21
\reference{} Hippelein, H., Maier, C., 
Meisenheimer, K., Wolf C., Fried, J.W., von Kuhlmann, B. , Kuemmel, M., 
Phleps, S., \& Roeser H.-J. 2003, \aa, 402, 65
\reference{} Isobe, T., Feigelson, E.~D., Akritas, M.~G., \& Babu, G.~J. 1990, ApJ, 364, 104
\reference{} Issa, M. R., MacLaren, I., \& Wolfendale, A.~W. 1990, A\&A, 236, 237
\reference{} Jansen, R.~A., Franx, M., Fabricant, D., \& Caldwell, N.
        2000a, ApJS, 126, 271 (NFGS)
\reference{} Kennicutt, R.~C. Jr. 1992, ApJS, 79, 255 
\reference{} Kennicutt, R. C., Bresolin, F., \& Garnett 2003, \apj, 591, 801
\reference{} Kewley, L.~J. \& Dopita, M.~A. 2003, ApJS, 142, 35 (KD03)
\reference{} Kewley, L. J., Dopita, M. A., Sutherland, R. S., 
Heisler, C. A. \& Trevena, J. 2001, \apj, 556, 121
\reference{} Kobulnicky, H.~A. \& Koo, D.~C. 2000, ApJ, 545, 712 (KK00)
\reference{} Kobulnicky, H.~A., Kennicutt, R.~C., \& Pizagno, J. 1998,
        ApJ, 514, 544
\reference{} Kobulnicky, H.~A. \& Phillips, A.~C. 2003, ApJ, 000 (KP03)
\reference{} Kobulnicky, H.~A. \& Zaritsky, D. 1999, ApJ, 511, 118 (KZ99)
\reference{} Kobulnicky, H.~A., Willmer, C.~N.~A., Weiner, B.~J., 
	Koo, D.~C., Phillips, A.~C., Faber, S.~M., Sarajedini, V.~L.,
	Simard, L., \& Vogt, N.~P. 2003, ApJ, 599, 1006 (Ke03)
\reference{} Kodama, T. \etal\ The Subaru/XMM-Newton Deep Survey Team, 2004, MNRAS, in press
\reference{} Lequeux, J., Peimbert, M., Rayo, J.~F., Serrano, A., \&    Torres--Peimbert, S. 1979, A\&A, 80, 155
\reference{} Lilly, S.~J., Le F\'evre, O., Crampton, D., Hammer, F., \& Tresse, L. 1995, ApJ, 455, 50
\reference{} Lilly, S.~M., Carollo, C.~M., \& Stockton, A.~N. 2003, ApJ, 597, 730 (LCS03)
\reference{} Leitherer, C. \etal\ 1999, ApJS, 123, 3 (Starburst99)
\reference{} Maciel, W.~J., Costa, R.~D.~D., Uchida, M.~M.~M. 2003, A\&A, 397, 667
\reference{} Maier, C., Meisenheimer, K., \& Hippelein, H. 2004, 418, 475
\reference{} McCall, M.~L., Rybski, P.~M., \& Shields, G.~A. 1985, 
	ApJS, 57, 1 (MRS)
\reference{} McGaugh, S. 1991, ApJ, 380, 140 (M91)
\reference{} McGaugh, S. 1998, private communication
\reference{} Mehlert, D. \etal\ 2002, A\&A, 393, 809
\reference{} Nagamine, K., Fukugita, \& Ostriker, Cen R. 2001, \apj, 558, 497
\reference{} Nagamine, K., Springel, V., \& Hernquist, L. 2004, \mnras, 348, 435
\reference{} Nishi, N, \& Tashiro, M. 2000, ApJ, 537, 50
\reference{} Pagel, B.~E.~J. Edmunds, M.~G., Blackwell, D.~E., Chun, 
		M.~S., \& Smith, G. 1979, MNRAS, 189, 95  
\reference{} Pettini, M., Shapley, A.~E., Steidel, C.~C., Cuby, J.-G.,
	Dickinson, M., Moorwood, A.~F.~M., Adelberger, K.~L., \& Giavalisco, M.
	2001, ApJ, 554, 981 (Pe01)
\reference{} Pr\'evot, M.~L., Lequeux, J., Maurice, E., Pr\'evot, L., \&
     Rocca-Volmerange, B. 1984, A\&A, 132, 389   
\reference{} Allende Prieto, C.~A., Lambert, D.~L., \& Asplund, M. 2001, ApJ, 556 L63
\reference{} Rowan-Robinson, M. 2001, \apj, 549, 745
\reference{} Savaglio, S., Glazebrook, K., Abraham, R.~G., Crampton, D., 
	Chen, H.-W., McCarthy, P.~J.~P.,\etal\ 2004, ApJ, in press
\reference{} Shapley, A. E., Ern, D. K., Pettini, M., Steidel, C. C., \&
Adelberger, K. L. 2004, \apj, {\it in press}, astro-ph/0405187 
\reference{} Shaver, P.~A., McGee, R.~X., Newton, L.~M., Danks, A.~C.,
	 Pottasch, S.~R. 1983, MNRAS, 204, 53
\reference{} Silva, L., Granato, G. L., Bressan, A., \& Danese, L. 1998, 
\apj, 509, 103
\reference{} Skillman, E.~D., Kennicutt, R.~C., \& Hodge, P. 1989, ApJ, 347, 875
\reference{} Skillman, E.~D., Tostoy, E., Cole, A.~A, Dolphin A.~E., Saha, A., Gallagher, 
	J.~S., Dohm-Palmer, R.~C., \& Mateo, M. 2003, ApJ, 596, 253
\reference{} Somerville, R.~S., Primack, J.~R., \& Faber, S.~M. 2001, MNRAS, 320, 504
\reference{} Stasi\'{n}ska, G. 2002, Rev. Mexicana Astron. Astrofis. Ser. Conf. 12, Ionized Gaseous Nebulae, ed. W. J. Henney \etal\ (Mexico, DF: UNAM), 62 
\reference{} Steidel, C., Shapley, A. E., Pettini, M., Adelberger, K. L., Erb, D., K., Reddy, M. A., \& Hunt, M. P. 2004, \apj, 604, 534 
\reference{} Sullivan, M., Treyer, M. A., Ellis, R. S., \& Mobasher, B. 2004, 
\mnras, 350, 21
\reference{} Tinsley, B.~M. 1974, ApJ, 192, 629 
\reference{} Tinsley, B.~M. 1980, Fundamentals of Cosmic Physics, 5, 287
\reference{} Vader, P. 1987, ApJ, 317, 128
\reference{} Walter, D.~K., Dufour, R.~J., \& Hester, J.~J. 1992, ApJ,
	397, 196
\reference{} Wheeler, J.~C., Sneden, C., \& Truran, J.~W. 1989, ARA\&A,
	27, 279
\reference{} Willmer, C.~N.~A. \etal\ 2004, ApJ, in prep
\reference{} Wirth, G.~D., Willmer, C.~N.~A., Amico. P. \etal\ 2004, ApJ, in prep
\reference{} Woosley, S. E. \& Weaver, T. A. 1995, ApJS, 101, 181
\reference{} Zaritsky, D., Kennicutt, R.~C., \& Huchra, J.~P. 1994, ApJ, 420,
	87 

%\reference{} McCall, M.~L., Rybski, P.~M., \& Shields, G.~A. 1985, 
%	ApJS, 57, 1 (MRS)


\end{references}
\end{document}